\documentclass[superscriptaddress,
showpacs,preprintnumbers,nofootinbib,eqsecnum,
amsmath,amssymb,aps,11pt, groupedaddress]{revtex4-2}
\usepackage[margin=1.0in]{geometry}
\usepackage{tikz,pgfplots}
\usepackage{amsmath,amssymb,bm}
\usepackage{graphics}
\usepackage{epstopdf}
\usepackage{subfigure}
\usepackage{braket}
\usepackage{dcolumn}
\usepackage{cancel}
\input{colordvi.tex}
\usepackage{longtable}
\usepackage{tabularx}
\usepackage{hyperref}
\usepackage{cleveref}
\usepackage{mathtools}
\usepackage[utf8]{inputenc}
\usepackage{xcolor}
\usepackage{soul}
\usepackage{pifont}
\definecolor{nicered}{rgb}{0,0.4,0.8}
\definecolor{nicegreen}{rgb}{.1,.5,.1}
\definecolor{darkblue}{rgb}{0,0,.8}
\hypersetup{colorlinks, citecolor=nicered ,linkcolor=darkblue, urlcolor=nicered}

\usepackage{multirow, makecell, cellspace, bigstrut}
\usepackage{pict2e}
\usepackage{booktabs}
\usepackage{diagbox}

\newcolumntype{C}{>{\centering\arraybackslash}X}

\begin{document}

\preprint{KIAS-Q26022} ~
\preprint{HRI-RECAPP-2026-09} ~ 
\preprint{CTPU-PTC-26-25}

\title{\large \bf 
Probe of Solar Neutrino Magnetic Moments through Spin–Flavor Precession: Resonance Structure and Antineutrino Appearance
}

\author{Pouya Bakhti$^1$}
\email{pouya\_bakhti@jbnu.ac.kr}

\author{Sudip Jana$^{2,3}$}
\email{sudip.jana@okstate.edu}

\author{Chui-Fan Kong$^4$}
\email{kongcf@ibs.re.kr}

\author{Seodong Shin$^{1,4,5}$}
\email{sshin@jbnu.ac.kr}

\author{Seokhoon Yun$^{6,4}$}
\email{seokhoon.yun@knu.ac.kr}

\affiliation{
$^1$Laboratory for Symmetry and Structure of the Universe,
Department of Physics, Jeonbuk National University, Jeonju, Jeonbuk 54896, Korea \\
$^2$ Harish-Chandra Research Institute, Chhatnag Road, Jhunsi, Prayagraj 211019, India\\
$^3$ Homi Bhabha National Institute, Training School Complex,
Anushakti Nagar, Mumbai 400 094, India \\
$^4$ Particle Theory and Cosmology Group, Center for Theoretical Physics of the Universe, Institute for Basic Science, Daejeon 34126, Korea \\
$^5$ School of Physics, Korea Institute for Advanced Study, Seoul 02455, Korea \\
$^6$ Department of Physics, Kyungpook National University, Daegu 41566, Korea 
}

\begin{abstract}

We investigate solar-neutrino spin--flavor precession (SFP) induced by magnetic moments in the three-active-flavor framework.
For Majorana neutrinos, SFP can convert solar neutrinos into antineutrinos of different active flavors.
In the Dirac case, SFP instead produces sterile right-handed states and can lead to the disappearance of active neutrinos. 
Using the full $6\times6$ Hamiltonians and GS98 and AGSS09 solar profiles, we examine propagation-eigenvalue crossings at $B_\perp=0$ and the projected magnetic couplings between the corresponding states. 
For normal mass ordering and $1\leq E_\nu/\mathrm{MeV}\leq20$, we confirm the absence of finite-density Majorana crossings. 
A magnetically coupled Dirac crossing emerges above approximately $12~\mathrm{MeV}$ but involves only a subdominant electron-flavor component, limiting resonant disappearance. 
Nonresonant Majorana conversion nevertheless offers a distinctive lepton-number-violating solar $\bar\nu_e$ signal, motivating our sensitivity study for the Jinping Neutrino Experiment. 
For a proposed $3~\mathrm{kt}$ detector operating for five to ten years, we project a 90\% C.L. sensitivity of $P(\nu_e\to\bar\nu_e)\simeq(0.85-1.3)\times10^{-5}$. 
In the $\mu_{12}$-only benchmark, optimistic solar-core transverse magnetic fields of $B_\perp=7-10~\mathrm{MG}$ imply a reach of $|\mu_{12}|\simeq(2.3-4.1)\times10^{-13}\,\mu_B$, numerically below existing direct-scattering limits and commonly quoted stellar-cooling bounds.
This could enable Jinping to provide one of the most stringent projected terrestrial sensitivities to Majorana transition magnetic moments.

\end{abstract}

\maketitle

\newpage


\newpage

\section{Introduction}
\label{sec:introduction}
Neutrinos are electrically neutral in the Standard Model (SM), but they can in principle acquire electromagnetic couplings to photons through quantum loop corrections. 
In the presence of new physics beyond the SM (BSM), guaranteed by the existence of neutrino oscillations, neutrino dipole moments are induced even solely from the neutrino mass term.
Due to the Lorentz structure, the 5-dimensional dipole moment operator requires the chirality flip of a neutrino, resulting in generic suppression by the neutrino mass term.
For instance, in the minimal Dirac neutrino extension of the SM with a right-handed singlet, the magnetic dipole moment is suppressed as \cite{Fujikawa:1980yx}
\begin{align}
    \mu_\nu \sim \frac{e G_F}{16\sqrt{2}\pi^2} m_\nu \sim 10^{-20} \left(\frac{m_\nu}{0.1\,{\rm eV}} \right) \mu_B\,,
    \label{eq:SMmdm}
\end{align}
where the Bohr magneton $\mu_B = (e/2m_e)\hbar \sim 5.788 \times 10^{-5}$ eV/T.
Taking a representative neutrino mass scale $m_\nu \lesssim 0.1$ eV, as motivated by cosmological constraints~\cite{Loureiro:2018pdz}, the neutrino magnetic dipole moment ($\nu$MDM) is expected to be below $10^{-20} \, \mu_B$.
In scenarios with Majorana neutrino masses, the expected transition magnetic moments resulting from SM interactions are further suppressed by a factor of $10^{-4}$ due to the unitarity of the neutrino mixing Pontecorvo-Maki-Nakagawa-Sakata (PMNS) matrix~\cite{Giunti:2014ixa}.

On the other hand, experimental and observational searches for $\nu$MDM have set bounds at the level of $\mathcal O (10^{-12} - 10^{-11}\,\mu_B)$, which are many orders of magnitude above the aforementioned minimal expectations~\cite{Giunti:2024gec}. 
Interestingly, dark matter direct detection experiments can be sensitive to $\nu$MDM because the magnetic moment contribution to solar neutrino scattering off target electrons or nuclei contains a term that scales as $d\sigma / d E_r \propto 1/E_r$.
This contribution is therefore enhanced at small recoil energies, $E_r \sim \mathcal O ({\rm keV})$. 
A stringent direct bound comes from elastic neutrino-electron scattering (E$\nu$ES), $\mu_\nu < 6.4 \times 10^{-12}\,\mu_B$~\cite{XENON:2022ltv}.
Solar neutrinos can also induce coherent elastic neutrino-nucleus scattering (CE$\nu$NS), as first observed in XENONnT~\cite{XENON:2024ijk} and PandaX-4T~\cite{PandaX:2024muv}. 
However, the SM contribution from $Z$ boson exchange, which is enhanced by the square of the neutron number, is much larger than the $\nu$MDM contribution, making the constraints from recent CE$\nu$NS observations about two orders of magnitude weaker~\cite{DeRomeri:2024hvc}. 
Reactor experiments provide another avenue for $\nu$MDM searches via neutrino scattering. 
The GEMMA spectrometer at the Kalinin nuclear power plant set an upper bound of $\mu_\nu < 2.9 \times 10^{-11}\,\mu_B$ by observing elastic scattering of $\bar \nu_e$ with electrons~\cite{Beda:2013mta}, whereas the CONUS experiment set an upper limit of $\mu_\nu < 7.5 \times 10^{-11}\,\mu_B$~\cite{CONUS:2022qbb}.

Astrophysical and cosmological observations can also probe $\nu$MDM indirectly.
Plasmon decay into a neutrino pair in stars provides an additional cooling channel.
Using observational data from the tip of the red giant branch (TRGB), Ref.~\cite{Capozzi:2020cbu} obtained the bound $\mu_\nu < 1.5 \times 10^{-12}\,\mu_B$.
For the Dirac neutrino scenarios, a right-handed neutrino can contribute to the effective number of relativistic species, $N_{\rm eff}$, which is the coefficient of the neutrino energy density relative to the photon density, i.e., $\rho_\nu \equiv N_{\rm eff} (7/8) (4/11)^{4/3} \rho_\gamma$.
Constraints from the cosmic microwave background (CMB) and big bang nucleosynthesis (BBN) on such right-handed neutrinos from the chirality flip of left-handed neutrinos in thermal plasma imply an indirect bound of $\mu_\nu < 2.7 \times 10^{-12}\,\mu_B$~\cite{Li:2022dkc}. However, in the Majorana scenario, cosmological constraints on the transition magnetic moments are considerably weaker, with upper limits of order $10^{-10}\,\mu_B$~\cite{Vassh:2015yza}.
Other searches and constraints are well summarized in Ref.~\cite{Giunti:2024gec}.

The large gap between the minimal neutrino-mass-induced prediction in 
Eq.~(\ref{eq:SMmdm}) and the existing experimental and observational limits leaves significant room for new physics contributions to the $\nu$MDM~\cite{Babu:2020ivd}. 
A precise study of $\nu$MDM is therefore essential, as it can serve as an indirect probe of well-motivated BSM scenarios. 
In the mass eigenstate basis, these BSM contributions are encoded in the coefficients $\mu_{ij}$ of the following effective $\nu$MDM operators:
\begin{align}
\mathcal L &\supset - \frac14 \sum_{i,j} \mu_{ij} \bar \nu_{i L}^c \sigma^{\mu \nu} \nu_{j L} F_{\mu \nu} ~ + ~ {\rm h.c.}\,,~~~~~{\rm Majorana} 
\label{eq:MDMMajorana}
\\
&\supset -\frac12 \sum_{i,j} \mu_{ij} \bar \nu_{i L} \sigma^{\mu \nu} \nu_{j R} F_{\mu \nu} ~ + ~ {\rm h.c.}\,,~~~~~{\rm Dirac}
\label{eq:MDMDirac}
\end{align}
where the additional factor of 1/2 in the Majorana case accounts for the double counting of identical fields.
For Majorana neutrinos, CPT invariance and self-conjugacy imply that the magnetic-moment matrix $\mu_M$ is antisymmetric in the mass eigenstate basis, $\mu_M^T=-\mu_M$. 
Hermiticity of the magnetic interaction requires $\mu_M^\dagger=\mu_M$, and together these conditions imply $\mu_M^*=-\mu_M$. 
Consequently, Majorana neutrinos possess only purely imaginary transition magnetic moments, with vanishing diagonal moments. 
For Dirac neutrinos, the magnetic-moment matrix satisfies $\mu_D^\dagger=\mu_D$ but need not be antisymmetric.

A magnetic field induces precession of a neutrino magnetic dipole moment, analogous to Larmor precession in quantum mechanics. 
Since the coefficient $\mu_{ij}$ can contain off-diagonal components in the mass basis, the precession of neutrinos happens not only in spin space but also in flavor space.~\footnote{Strictly speaking, the precession changes the mass eigenstate; the corresponding flavor change follows from neutrino mixing.}
This phenomenon is referred to as the spin-flavor precession (SFP).

Solar neutrinos have long been regarded as a suitable probe of the SFP, since the component of the solar magnetic field transverse to the neutrino propagation direction, $B_\perp$, can induce neutrino spin precession.~\footnote{Helioseismic studies expect that $B_\perp$ can reach up to 7 MG in the solar core ($< 0.2\,R_\odot$), which includes the $^8$B neutrino production region ($\sim 0.05\,R_\odot$)~\cite{Antia:2008}.
By contrast, studies of the radiation zone and tachocline suggest values smaller than 600 G for stability~\cite{Kitchatinov:2008}.
}
Historically, SFP of solar neutrinos was proposed as a possible solution to the solar neutrino problem~\cite{Cisneros:1970nq}. 
In the Dirac neutrino case, $\nu_{eL}$ can be converted into a right-handed state, $\nu_R$, which is sterile under the weak interaction and therefore remains undetected.
Motivated by this possibility, periodic modulation of the solar magnetic field was proposed as a possible explanation for decade-scale variations in the observed solar neutrino flux~\cite{Voloshin:1986ty,Okun:1986na}.

In the presence of matter, however, $\nu_{eL}$ and $\nu_{eR}$ become non-degenerate because only $\nu_{eL}$ receives the weak matter potential associated with the Mikheyev-Smirnov-Wolfenstein (MSW) matter effect~\cite{Wolfenstein:1977ue,Mikheyev:1985zog}.
Thus the resonance condition of the same flavor conversion $\nu_{eL} \to \nu_{eR}$ requires a very high solar neutron density compared to electrons, $N_n = 2 N_e$ or equivalently $Y_e = N_e/(N_e + N_n) = 1/3$ in the simplified 2-flavor scheme.
This neutron-rich composition is not realized at finite density in standard solar models.
Thus $\nu_{eL}$ is just reflipped into itself; for a localized, non-twisting magnetic field that turns on and off slowly, an adiabatically followed eigenstate therefore starts as $\nu_{eL}$, becomes mixed inside the field region, and returns to $\nu_{eL}$ when the field vanishes, suppressing the final $\nu_{eR}$ population~\cite{Akhmedov:1988uk}.
On the other hand, spin-flavor transitions induced by transition magnetic moments, $\mu_{ij}$ where $i \ne j$, have the potential to undergo resonant enhancement analogous to the MSW effect during the neutrino propagation through the solar medium.
This occurs when the kinetic energy difference, $\Delta m_{ij}^2/2E$, is nearly canceled by the matter-induced potential energy difference with a tiny correction from the off-diagonal term proportional to $\mu_{ij} B_\perp$~\cite{Lim:1987tk,Akhmedov:1988uk}.
This phenomenon, subsequently known as resonant spin-flavor precession (RSFP), received considerable attention in the late 1980s and early 1990s as a possible mechanism for explaining the solar neutrino problem ~\cite{Lim:1987tk,Akhmedov:1988uk}.

The solar neutrino problem was eventually understood as a consequence of neutrino flavor oscillations~\cite{SNO:2001kpb,SNO:2002tuh,Super-Kamiokande:1998kpq,Super-Kamiokande:2001ljr,KamLAND:2013rgu}, with matter effects playing an important role in the conversion of neutrinos inside the Sun \cite{Wolfenstein:1977ue, Mikheev:1986gs, Mikheyev:1985zog}. Nevertheless, the possibility of subleading contributions from SFP remains viable, offering a complementary avenue for probing BSM physics.
Future precision measurements of the solar neutrino flux may reach the sensitivity required to probe SFP effects.~\footnote{Precision measurements of solar-neutrino oscillation parameters using reactor experiments such as JUNO and $\nu$EYE \cite{Seo:2023xku, NuEYE:2026gyx} at Yemilab, together with solar-neutrino observations, open a new window for probing BSM scenarios~\cite{Bakhti:2023vzn}.}
Two qualitatively distinct scenarios should be considered.
If neutrinos are Majorana particles, only transition magnetic moments are allowed due to the CPT invariance,
and hence the SFP can convert $\nu_e$ into active antineutrinos of different flavors.
These antineutrinos subsequently oscillate into electron antineutrinos during propagation to the Earth, leading to a possible $\bar\nu_e$ appearance signal detectable via inverse beta decay (IBD) in neutrino experiments~\cite{Akhmedov:1991nt,Raghavan:1991em}.
If neutrinos are Dirac particles, magnetic interactions flip chirality,
producing sterile right-handed states that do not participate in weak interactions,
leading instead to an apparent disappearance or spectral suppression of active solar neutrinos~\cite{Akhmedov:1987nc,Lim:1987tk,Akhmedov:1991nt}.
Thus, the Majorana and Dirac cases lead to markedly different observational signatures.

In this paper, we revisit solar-neutrino SFP in the complete three-active-flavor framework for both Majorana and Dirac neutrinos, assuming normal mass ordering and a single independent mass-basis transition moment $\mu_{12}$. 
We construct the corresponding $6 \times 6$ evolution Hamiltonians and reassess the resonance condition in this multilevel system. 
Rather than relying on equalities between basis-dependent diagonal elements, we identify a candidate RSFP as a crossing of the $B_\perp=0$ propagation eigenvalues connected by a nonzero projected magnetic coupling; a finite magnetic field turns such a crossing into an avoided crossing. 
Using updated oscillation parameters and the GS98 and AGSS09 solar profiles, we find no finite-density Majorana crossing for $1\leq E_\nu/\mathrm{MeV}\leq20$, so Majorana SFP is nonresonant within our setup. 
Nevertheless, a potentially observable lepton-number-violating solar $\bar \nu_e$ appearance signal may still arise through nonresonant SFP.
In the Dirac case, a magnetically coupled crossing, $\lambda_1^L=d_1$, appears above approximately $12~\mathrm{MeV}$. 
However, the participating left-handed branch has only a small electron-flavor component, and the crossing lies deep in the solar core with high electron number density. 
We therefore discuss the possible Dirac disappearance only briefly and focus our experimental analysis on the distinctive Majorana solar $\bar \nu_e$ appearance channel.

For this channel, we assess the prospective sensitivity of the proposed Jinping Neutrino Experiment through inverse beta decay. 
For a $3~\mathrm{kt}$ fiducial target and five to ten years of exposure, we project a $90\%$ C.L. sensitivity of $P(\nu_e\to\bar \nu_e)\simeq(0.85 - 1.3)\times 10^{-5}$. 
For benchmark solar-core fields of $B_\perp=7 - 10~\mathrm{MG}$, this corresponds to $|\mu_{12}|\simeq(2.3 - 4.1) \times 10^{-13}\,\mu_B$. 
The remainder of this paper is organized as follows. Section~\ref{sec:theory} introduces the evolution Hamiltonians for both Majorana and Dirac neutrinos, together with the conventions adopted for the neutrino magnetic moments. Section~\ref{sec:resonance} investigates the resonance structure of neutrino propagation using the GS98 and AGSS09 solar density profiles. Section~\ref{sec:appearance} presents our analysis of solar antineutrino appearance and the resulting sensitivity projected for the Jinping experiment. Finally, Section~\ref{sec:conclusions} summarizes our main findings and conclusions.

\section{Theoretical framework}
\label{sec:theory}

In this section, we formulate the neutrino propagation in the Sun including both standard flavor oscillations in matter and SFP induced by a transverse magnetic field.
In vacuum, the flavor-basis neutrino amplitudes $\nu_{\alpha L}$ are related to the mass-basis neutrino amplitudes $\nu_{i L}$ through the PMNS matrix $U$ as
\begin{align}
\nu_{\alpha L} = \sum_i U_{\alpha i} \nu_{i L} \,,
\end{align}
where $\nu_{\alpha L}$ is a component of the column vector $\nu_{f L} = (\nu_{e L}, \nu_{\mu L}, \nu_{\tau L})^T $.

In the presence of matter, the coherent forward scattering of neutrinos with the materials in the medium is affected by the matter potentials $V_e^{\rm CC} = \sqrt 2 G_F N_e (\vec r)$ arising from the charged-current (CC) scattering with electrons and $V_n^{\rm NC} = - \sqrt 2 G_F N_n (\vec r) / 2$ by the neutral-current (NC) scattering with neutrons, where $G_F$ is the Fermi constant, $N_e(\vec r)$ and $N_n(\vec r)$ are the electron and neutron number densities, respectively, at the position $\vec r$ from the solar core.
Here we assume the matter through which the neutrino propagates is electrically
neutral. 
Therefore, the contributions of electron and proton, with almost opposite weak vector charges, to NC scattering will cancel out. 
The overall matter potentials of $\nu_{eL}$, $\nu_{\mu L}$, and $\nu_{\tau L}$ are 
\begin{align}
V_e &= V_e^{\rm CC} + V_n^{\rm NC} = \sqrt{2}G_F\left(N_e-\frac{1}{2}N_n\right)\,,~V_\mu = V_\tau = V_n^{\rm NC} = -\sqrt{2}G_F\frac{1}{2}N_n\,,
\end{align}
while the antineutrinos feel the opposite potential $V_{\bar \alpha} = - V_\alpha$ since the vector current is odd under charge conjugation.
Given the above matter potentials, the coherent evolution of the flavor amplitudes is described in the form of a Schr\"odinger-like equation:
\begin{align}
    i \frac{d}{d L} \nu_{f L} = H \nu_{f L} &=  \left[\frac{1}{2E} U M^2 U^\dagger + V_m \right] \nu_{f L}\,,
    \label{eq:MatterH}
\end{align}
where $M^2 = {\rm diag}(0, \Delta m_{21}^2, \Delta m_{31}^2)$ and $V_m = {\rm diag}(V_e^{\rm CC}, 0, 0)$ for relativistic neutrinos with baseline $L$ in natural units $c = \hbar = 1$.
The universal diagonal contribution $V_n^{\rm NC}$, which is irrelevant for the flavor transition, is omitted from $V_m$.
Due to the hierarchy between the two mass splittings, $\Delta m_{21}^2 \ll |\Delta m_{31}^2|$, solar neutrino conversion is governed primarily by $\theta_{12}^m$ where approximately
\begin{align}
\tan2\theta_{12}^m = \frac{(\Delta m_{21}^2/2E) \sin 2\theta_{12}}{(\Delta m_{21}^2 /2E) \cos2\theta_{12} - V_e^{\rm CC} \cos^2\theta_{13}}\,,
\end{align}
for $V_e^{\rm CC} \ll |\Delta m_{31}^2| / 2E$ with the 3rd state being nearly decoupled and hence $\theta_{13}^m \approx \theta_{13}$.

\subsection{Majorana case}

In the presence of SFP, the Hamiltonian is extended to a $6 \times 6$ matrix for the 3-active neutrino flavor scenarios by the addition of the off-diagonal contributions from the $\nu$MDM. 
In the 3-active Majorana neutrino scenarios, we use the 6-dimensional basis
\begin{align}
  \Psi_M^{3\nu}=(\nu_{eL},\nu_{\mu L},\nu_{\tau L},
  \nu_{eL}^c, \nu_{\mu L}^c, \nu_{\tau L}^c)^T\,,
\end{align}
and the Hamiltonian is
\begin{align}
  H_M^{3\nu}=
  \begin{pmatrix}
  H_\nu & \mu_M^\dagger B_\perp \\
  \mu_M B_\perp & H_{\bar\nu}
  \end{pmatrix}
  =
  \begin{pmatrix}
  H_\nu & - \mu_M^\ast B_\perp \\
  \mu_M B_\perp & H_{\bar\nu}
  \end{pmatrix}
  \,,
  \label{eq:H_M_3nu}
\end{align}
where
\begin{align}
  H_\nu &= \frac{1}{2E}U M^2 U^\dagger+\mathrm{diag}(V_e,V_\mu,V_\tau)\,, \\
  H_{\bar\nu} &= \frac{1}{2E}U^* M^2 U^T-\mathrm{diag}(V_e,V_\mu,V_\tau)\,,
\end{align}
for the PMNS matrix $U$ including the diagonal Majorana phase matrix diag(1,$e^{i \alpha_1}, e^{i\alpha_2}$).
The magnetic moment matrix of Majorana neutrinos is antisymmetric, $\mu_M^T = - \mu_M$, and hence the diagonal magnetic moments vanish and $\mu_M^\dagger = (\mu_M^T)^\ast = - \mu_M^\ast$, as mentioned in Sec.~\ref{sec:introduction}.
In general, the transverse magnetic field can be twisting by being complex, $B_\perp e^{i \phi}$, providing an extra diagonal term in the reference frame rotating with the angular velocity of the magnetic field~\cite{Akhmedov:1993sh}. 
For simplicity, however, we only consider a non-twisting magnetic field here, i.e., $\phi = 0$. 
Note that the $\nu$MDM in the flavor basis $\mu_{M \alpha \beta}$, where $\alpha$ is the flavor index of neutrinos and $\beta$ is that of antineutrinos, is related to the $\nu$MDM in the mass eigenstate basis $\mu_{ij}$ defined in Eq.~(\ref{eq:MDMMajorana}) as
\begin{align}
    \mu_{M \alpha \beta} = \sum_{i,j = 1}^3 U^\ast_{\alpha i} \mu_{ij} U^\dagger_{j \beta}\,,
    \label{eq:MDMfM}
\end{align}
since $\nu_{\alpha L}^c = \sum_i U^\ast_{\alpha i} \nu_{i L}^c$, so that $\bar \nu_{\alpha L}^c$ and the $\nu$MDM in the flavor eigenstate basis are in the operator 
\begin{align}
\mathcal L \supset -\frac14 \sum_{\alpha, \beta} \mu_{M \alpha \beta} \bar \nu_{\alpha L}^c \sigma^{\mu \nu} \nu_{\beta L} F_{\mu \nu} ~ + ~ {\rm h.c.}\,.
\end{align}
Due to the anti-symmetric feature of $\mu_M$, direct conversion of $\nu_e \to \bar \nu_e$ is forbidden.
Hence the transition occurs via the two-step processes of SFP and oscillation: $\nu_e \to \bar \nu_\mu \to \bar \nu_e$ or $\nu_e \to \nu_\mu \to \bar \nu_e$ in the two flavor assumption~\cite{Akhmedov:1991nt}.
Again, the anti-symmetry of $\mu_M$ suppresses the simultaneous occurrence of the two processes inside the Sun.
Since the magnetic field outside the Sun is relatively weak, the former process is preferred: the SFP transition of $\nu_e \to \bar \nu_\mu$ occurs first inside the Sun, followed by the oscillation $\bar \nu_\mu \to \bar \nu_e$ while propagating from the Sun to the Earth.
Note that the MSW matter effect also operates during neutrino propagation inside the Sun.
In the three-active-flavor Majorana framework considered here, SFP produces a subdominant correction to standard MSW evolution in the Sun, owing to the absence of an SFP resonance and the small magnetic coupling $|\mu_{12}B_\perp|$ in our benchmark scenarios.
More details on the resonance conditions will be discussed later.

\subsection{Dirac case}

For three active Dirac neutrinos, the spin-flavor basis is
\begin{align}
  \Psi_D^{3\nu}=(\nu_{eL},\nu_{\mu L},\nu_{\tau L},
  \nu_{eR},\nu_{\mu R},\nu_{\tau R})^T \,.
\end{align}
The Hamiltonian is
\begin{align}
  H_D^{3\nu}=
  \begin{pmatrix}
  H_L & \mu_D B_\perp \\
  \mu_D^\dagger B_\perp & H_R
  \end{pmatrix}
  \label{eq:H_D_3nu}
\end{align}
with
\begin{align}
  H_L &= \frac{1}{2E}U_L M^2 U_L^\dagger+\mathrm{diag}(V_e,V_\mu,V_\tau)\,, \\
  H_R &= \frac{1}{2E}U_R M^2 U_R^\dagger \,,
\end{align}
where $\nu_{\alpha L} = \sum_i U_{L \alpha i} \nu_{i L}$ while $\nu_{\alpha R} = \sum_i U_{R \alpha i} \nu_{i R}$ for Dirac neutrinos.
Unlike $U_L$, we have freedom to choose the matrix $U_R$ since it is not determined by the neutrino oscillation experiments. 
Following the convention of the quark mixing matrix, the matrix $U_R$ is usually set to be an identity matrix, while $U_L$ is the PMNS matrix with the Dirac CP phase only.
The right-handed states have no weak matter potential since they are sterile to the weak interactions.
Here $\mu_D$ is a general Dirac magnetic-moment matrix with the relation
\begin{align}
    \mu_{D \alpha \beta} = \sum_{i,j = 1}^3 U_{L\,\alpha i}\mu_{ij} U_{R\, j \beta}^\dagger \,.
    \label{eq:MDMfD}
\end{align}

For the simplest case in which only one $\nu$MDM is non-zero for solar neutrinos - here we choose $\mu_{12} \ne 0$ - it is convenient to formulate the scenario in terms of the instantaneous matter eigenvalues and the transformed $\nu$MDM matrix in the primed flavor basis, as defined in Ref.~\cite{Akhmedov:2022txm}. In the primed flavor basis, the left-handed fields are related to the standard flavor basis via $\nu_{\alpha L} = O_{23}\Gamma_\delta O_{13} \nu'_{\alpha L}$, where $O_{ij}$ is the orthogonal rotation matrix in the $i\text{-}j$ plane and $\Gamma_\delta = {\rm diag}(1,1,e^{i \delta_{\rm CP}})$ contains the Dirac CP-violating phase. This transformation isolates the dynamics of the first and second mass/matter eigenstates from the atmospheric sector without altering physical observables or resonance conditions.
By removing the atmospheric (2-3) mixing and the (1-3) CP-phase rotation, the active neutrino propagation Hamiltonian assumes a nearly block-diagonal structure where the dominant solar dynamics are confined to the (1-2) sector. 
The matter-induced coupling between the first and third states is strictly proportional to $s_{13}c_{13}V_e^{\rm CC}$.~\footnote{See Eq. (2.6) of Ref.~\cite{Akhmedov:2022txm}.} Because this term is extremely small compared to the atmospheric-scale kinematic splitting $\vert{}\Delta m_{31}^2\vert{}/2E$, the third state effectively decouples from the solar evolution. Consequently, the active neutrino propagation is reduced to a two-level system, allowing for simple analytical expressions for the instantaneous matter eigenvalues, the effective mixing angle, and the associated propagation phases.
In the Majorana neutrino scenario, $\mu_{e' \mu'}$ and $\mu_{12}$ are aligned as $\mu_{e' \mu'} = \mu_{12} e^{-i \alpha_1}$ while $\mu_{\mu' \tau'}$ and $\mu_{e' \tau'}$ have no dependence on $\mu_{12}$ in this basis.
Then the dominant contribution to $P(\nu_e \to \bar \nu_e)$ is proportional to $|\mu_{12}|^2$ in reasonable approximations~\cite{Akhmedov:2022txm}, which makes our simple analysis assuming only $\mu_{12} \ne 0$ efficient.
In the Dirac neutrino scenario, the primed basis in the left-handed sector isolates the ordinary solar (1-2) dynamics in $H_L$.
But rotating the right-handed sector to the primed basis as in the left-handed sector makes $H_R$ non-diagonal and hence it is better to keep the conventional $\nu_R$ basis.

\section{Discussion on resonance conditions}
\label{sec:resonance}

In the presence of magnetic fields and $\nu$MDM, solar neutrino propagation is governed simultaneously by the standard MSW matter effect and SFP, requiring some delicate treatments of the resonance conditions~\cite{Akhmedov:1993sh,Friedland:2005xh,Joshi:2019dcj}.
The RSFP condition can be commonly characterized at three related levels: (i) equality of selected diagonal Hamiltonian elements, $H^\nu_{aa} = H^{\bar \nu}_{bb}$ or $H^L_{aa} = H^R_{bb}$, which provides a basis-dependent nominal two-state criterion; (ii) crossing of the corresponding propagation eigenvalues in the absence of the magnetic coupling; and (iii) at finite $B_\perp$, an avoided crossing characterized by a local minimum of the full eigenvalue splitting accompanied by an exchange of the corresponding eigenstate compositions.
When the ordinary off-diagonal mixing terms are small compared with the relevant diagonal splittings, the chosen flavor or primed basis approximately coincides with the propagation-eigenstate basis, so that criteria (i) and (ii) become nearly equivalent.
If, in addition, the magnetic coupling is sufficiently weak and slowly varying across the crossing region, the center of the avoided crossing in criterion (iii) also approximately coincides with them.
In a general multilevel system, however, these three criteria need not be identical: equality of two selected diagonal entries is an exact resonance condition only for an isolated two-state system, whereas the propagation-eigenvalue crossing provides a {\it basis-independent} diagnostic of the relevant magnetic transition.

At fixed neutrino energy and solar radius, we can write the reduced Hamiltonian
for $N=3$ flavor coordinates as
\begin{align}
 \mathcal H_X^{(N)}=
 \begin{pmatrix}H_L&\mathcal B_X\\ \mathcal B_X^\dagger&H_X\end{pmatrix},
 \label{eq:insert-full-H}
\end{align}
where $X=\bar\nu$ (Majorana) or $R$ (Dirac) and
$\mathcal B_X\propto\mu_XB_\perp$. 
Following the category (ii) discussed above, we omit $\mathcal B_X$ when
diagonalizing each propagation block, while retaining all ordinary
vacuum and matter mixing inside $H_L$ and $H_X$, and then project the
physical magnetic block into that basis:
\begin{align}
 W_L^\dagger H_LW_L&=\operatorname{diag}(\lambda_i^L)\,,&
 W_X^\dagger H_XW_X&=\operatorname{diag}(\lambda_j^X)\,,\nonumber\\
 g_{ij}^X&=(W_L^\dagger\mathcal B_XW_X)_{ij}\,,&
 \Delta_{ij}^X&=\lambda_i^L-\lambda_j^X \,,
 \label{eq:insert-levels-coupling}
\end{align}
where $W_{L,X}$ is the matrix diagonalizing the $3 \times 3$ block matrix $H_{L,X}$ and $g_{ij}^X$ are the off-diagonal $3 \times 3$ block in the new basis.
Here $i,j=1,\ldots,N$.  
For a pair isolated from the other $2N-2$ levels,
\begin{align}
 \mathcal H_{ij}^{\rm loc} = \begin{pmatrix}\lambda_i^L&g_{ij}^X\\g_{ij}^{X*}&\lambda_j^X\end{pmatrix}\,,
 \qquad
 \sin^22\Theta_{ij}^X=
 \frac{4|g_{ij}^X|^2}{(\Delta_{ij}^X)^2+4|g_{ij}^X|^2}\,.
 \label{eq:insert-pair-angle}
\end{align}
Thus $\Delta_{ij}^X=0$ is a basis-independent magnetic-off crossing
candidate inducing the maximal mixing $\Theta_{ij}^X = \pi / 4$, provided $g_{ij}^X\ne0$.  
At finite $B_\perp$, the physical signature
is an avoided crossing together with an exchange of branch projectors
or channel compositions. 
The isolated two-level treatment may, however, become inadequate when distinct crossing regions occur sufficiently close to one another.
In particular, if their spatial separation becomes comparable to the relevant oscillation length, or if their resonance widths overlap, coherent oscillations and interference between successive transitions may become important, and the crossings cannot generally be treated independently.
If additional propagation levels participate in such nearby crossings, the complete $2N\times2N$ evolution should be followed.
A rapid spatial variation of $B_\perp(r)$, or more generally of the projected magnetic coupling $g_{ij}^X(r)$, can also invalidate the approximation that the coupling remains nearly constant across the resonance width and can significantly affect the adiabaticity of the transition.
The historical solar-sector condition for the 3-flavor Majorana case for example
\begin{align}
H^\nu_{e'e'}=H^{\bar\nu}_{\mu'\mu'}
 \quad\Longleftrightarrow\quad
 c_{13}^2V_{\rm CC}+2V_n
 =\frac{\Delta m_{21}^2}{2E_\nu}\cos2\theta_{12}
 \label{eq:insert-nominal-line}
\end{align}
is therefore the \emph{nominal} primed/effective-two-flavor layer, not itself a basis-invariant magnetic-off crossing.

\subsection{Majorana neutrino resonance}

\begin{figure}
    \centering
    \includegraphics[width=\linewidth]{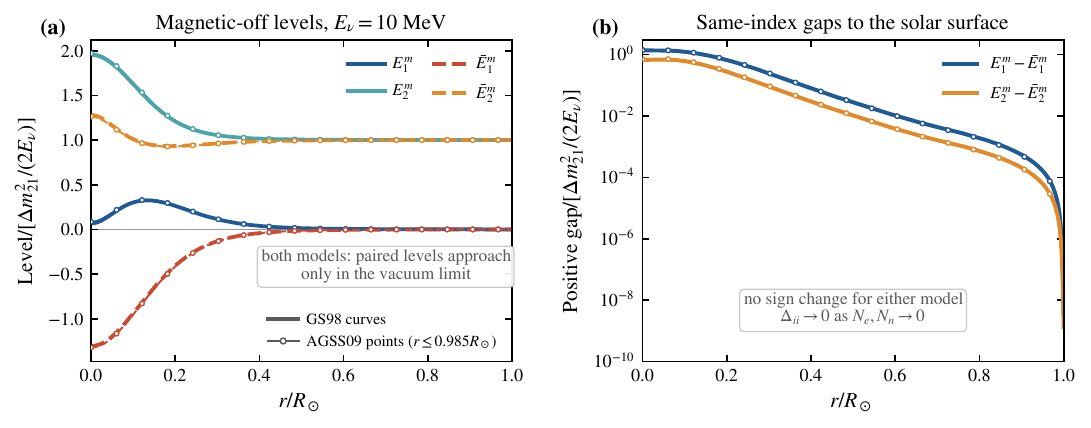}
    \caption{
Propagation-level diagnostics in realistic solar matter for the 3-flavor Majorana scenario.
\textbf{(a)} The two lowest neutrino and antineutrino matter eigenvalues at $E_\nu=10~\mathrm{MeV}$, $E_i^m(r;E_\nu)\equiv\lambda_i[H_\nu^f(r;E_\nu)$] and 
$\bar E_j^m(r;E_\nu)\equiv\lambda_j[H_{\bar\nu}^f(r;E_\nu)$],
normalized by $\Delta_{21}\equiv\Delta m_{21}^2/(2E_\nu)$.  
Continuous curves show the GS98 result, while open markers show AGSS09, which is displayed only up to its tabulated endpoint $r=0.985R_\odot$ and is not extrapolated.
\textbf{(b)} Positive same-index gaps $E_1^m-\bar E_1^m$ and $E_2^m-\bar E_2^m$ on a logarithmic scale. Neither gap changes sign at finite solar density, and both approach zero only in the vacuum limit $N_e,N_n\rightarrow0$. 
    }
    \label{fig:majorana-diagnostic}
\end{figure}

Figure~\ref{fig:majorana-diagnostic} gives the resulting diagnostic
for normal ordering in the 3-flavor Majorana scenario, with $\sin^2\theta_{12}=0.3088$, $\sin^2\theta_{13}=0.02248$,
$\Delta m_{21}^2=7.537 \times10^{-5}\,\mathrm{eV}^2$, and
$\Delta m_{31}^2=2.511 \times10^{-3}\,\mathrm{eV}^2$ from NuFIT 6.1 (2025)~\cite{Esteban:2024eli}~\footnote{The website is www.nu-fit.org.} at $E_\nu = 10$ MeV as a representative energy of ${}^8$B neutrino benchmark, consistent with the rest of our solar-SFP analysis. 
The vertical axis of panel (a) corresponds to the dimensionless eigenvalues normalized by  $\Delta_{21} \equiv \Delta m_{21}^2 / 2 E_\nu$, while that of (b) is the absolute values of the eigenvalue gaps.
The realistic GS98 and
AGSS09 number densities are reconstructed from the tabulated solar-model
abundances.
Since the eigenvalues of the reduced evolution Hamiltonian are obtained after removal of the common relativistic energy, $p + m_1^2 / 2E_\nu$, the negative red dashed branch is $\bar E_1^m$ and does not represent a negative physical energy; $\bar E_2^m$ remains positive. 
For $r \gtrsim 0.4 R_\odot$, the eigenvalues $E_1^m$ ($E_2^m$) and $\bar E_1^m$ ($\bar E_2^m$) approach each other. 
However, as shown in panel (b), neither gap changes sign at finite solar density, and both approach zero only in the vacuum limit $N_e\,,N_n \rightarrow 0$, where the corresponding neutrino and antineutrino vacuum levels become trivially degenerate.
This vacuum degeneracy does not constitute an RSFP, since the corresponding same-index Majorana magnetic couplings vanish in the vacuum mass basis.

Performing an independent scan of all nine neutrino–antineutrino gaps over $1\le E_\nu/\mathrm{MeV}\le20$,
we find none of the nine $E_i^m-\bar E_j^m$ gaps has a finite-density zero, i.e., no magnetic-off crossing for either solar model over $1\leq E_\nu/\mathrm{MeV}\leq20$. 
The nominal line in
Eq.~\eqref{eq:insert-nominal-line} does occur above approximately $3.70$ MeV
(GS98) or $3.60$ MeV (AGSS09), but it is not a physical resonance from the basis-independent magnetic-off crossing.
Solar $\bar\nu_e$ appearance can therefore still be generated by
\emph{non-resonant} SFP.  For a smooth non- or slowly twisting weak field, the
perturbative three-flavor amplitude is correspondingly endpoint dominated~\cite{Akhmedov:2022txm}.

\subsection{Dirac neutrino resonance}

\begin{figure}
    \centering
    \includegraphics[width=\linewidth]{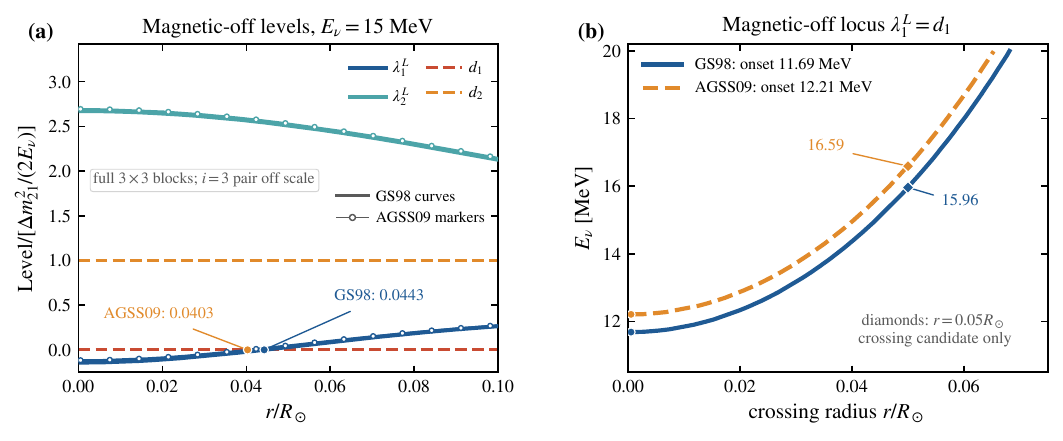}
    \caption{
Magnetic-off eigenvalue diagnostic for three-flavor Dirac neutrinos.
\textbf{(a)} The two lowest left-handed matter eigenvalues $\lambda_{1,2}^L$ and right-handed levels $d_{1,2}$ at $E_\nu=15~\mathrm{MeV}$, normalized by $\Delta_{21}=\Delta m_{21}^2/(2E_\nu)$. Continuous curves show GS98, while open markers show AGSS09. The equality $\lambda_1^L=d_1$ occurs at $r\simeq0.044R_\odot$ for GS98 and $r\simeq0.040R_\odot$ for AGSS09. The third pair is omitted because it lies outside the displayed vertical range.
\textbf{(b)} Locus of the magnetic-off equality $\lambda_1^L(r;E_\nu)=d_1$ in the $(r,E_\nu)$ plane. A finite-density solution first appears at $E_\nu\simeq11.69~\mathrm{MeV}$ for GS98 and $12.21~\mathrm{MeV}$ for AGSS09. The diamonds indicate the energies at which the equality occurs at the representative production radius $r_0=0.05R_\odot$: $15.96~\mathrm{MeV}$ and $16.59~\mathrm{MeV}$, respectively. Consequently, a neutrino produced at $r_0=0.05R_\odot$ and propagating outward encounters this layer only above the corresponding energy. 
    }
    \label{fig:dirac-diagnostic}
\end{figure}

The three-flavor Dirac case is different because the sterile right-handed
block has eigenvalues $d_j \equiv \Delta m_{j1}^2/(2E_\nu)$, while the left-handed
active block retains the neutral-current shift relative to those sterile
states.  
In Fig.~\ref{fig:dirac-diagnostic}, we show the  magnetic-off eigenvalue diagnostic for three-flavor Dirac neutrinos.
In panel (a), we fix the neutrino energy $E_\nu = 15$ MeV, to make the level crossing visible.
As seen in panel (b), an exact $\lambda_1^L=d_1$ crossing first appears at approximately
$11.69$ MeV (GS98) or $12.21$ MeV (AGSS09); at $15$ MeV it is located at
$r\simeq0.044R_\odot$ or $0.040R_\odot$, respectively. 
The third pair is omitted because it lies outside the displayed vertical range.
For Dirac mass eigenstates whose masses have been chosen real and positive, Hermiticity of the electromagnetic current implies that the single-transition benchmark obeys
$(\mu_D)_{12}=(\mu_D)_{21}^*$, and has nonzero projected coupling to this
crossing for the benchmark considered.  

The condition $\lambda_1^L = d_1$ identifies a physically coupled Dirac RSFP between $\nu_{1m}^L$ and the sterile state $\nu_{1R}$ when only $\mu_{12} B_\perp = \mu_{21}^\ast B_\perp \ne 0$ since $g_{11}^D = B_\perp (W_L^\dagger \mu_D)_{11} = B_\perp (W_L^\dagger U_L)_{12} \, \mu_{21}$ in the widely used convention $U_R = W_R = 1_{3 \times 3}$.
Nevertheless, this crossing is expected to have only a limited impact on the observable solar $\nu_e$ flux, because the participating $\lambda_1^L$ branch
has only a small electron-flavor component in the relevant region and the
crossing occurs deep in the solar core, inside the production radius of a
substantial fraction of the solar neutrinos.
For example, at the GS98 15 MeV crossing, $|\langle \nu_e | \nu_{1L}^m\rangle|^2 \simeq 0.034$, while $|\langle \nu_e | \nu_{2L}^m\rangle|^2 \simeq 0.940$.

Our brief propagation calculation predicts a spectral distortion in $P_{ee}\equiv P(\nu_e\to\nu_e)$ that grows from the sub-percent level near $E_\nu\simeq12~\mathrm{MeV}$ to approximately 2\% near 15 MeV. 
Observing this feature would, however, be extremely challenging because of the limited endpoint statistics, uncertainties in the normalized high-energy $^{8}$B spectral shape, the growing $hep$-neutrino contribution and its normalization uncertainty, and detector-response systematics. 
We therefore regard this downturn primarily as a characteristic theoretical signature and do not pursue a dedicated detector-level sensitivity analysis here. 
Future high-statistics measurements at Hyper-Kamiokande (HK) and DUNE, combined with improved determinations of the $^{8}$B and $hep$ spectra and detector responses, could nevertheless constrain larger distortions or the corresponding Dirac magnetic-moment parameter space.

On the other hand, it would be relatively easy to satisfy the resonance conditions for the hypothetical (3+1)-flavor scenarios due to the existence of an additional parameter $\Delta m_{01}^2$ where $m_0$ is the mass of the 4th generation sterile neutrino~\cite{Picariello:2007qj,Das:2009kw}.
We leave more details in this direction to future work.

\section{Solar antineutrino appearance analysis}
\label{sec:appearance}

In this section, we discuss the analysis details of the solar antineutrino appearance channel and show the experimental sensitivities.
Because of the limited energy of solar neutrinos, the only antineutrino appearance channel we can observe is $\nu_e\rightarrow \bar\nu_e$ which can be probed via IBD.
Among solar neutrino experiments, we expect Jinping Neutrino Experiment (Jinping), a proposed observatory in China Jinping Underground Laboratory (CJPL),
has a promising capability of searching for solar antineutrino appearance due to its relatively small background contamination from reactor neutrinos, compared to other experiments such as JUNO~\cite{JUNO:2015zny}, Super-Kamiokande (SK)~\cite{Super-Kamiokande:2002weg}, and HK~\cite{Hyper-Kamiokande:2018ofw}.

\subsection{Signal and background estimation}

\begin{figure}
    \centering    \includegraphics[width=0.5\linewidth]{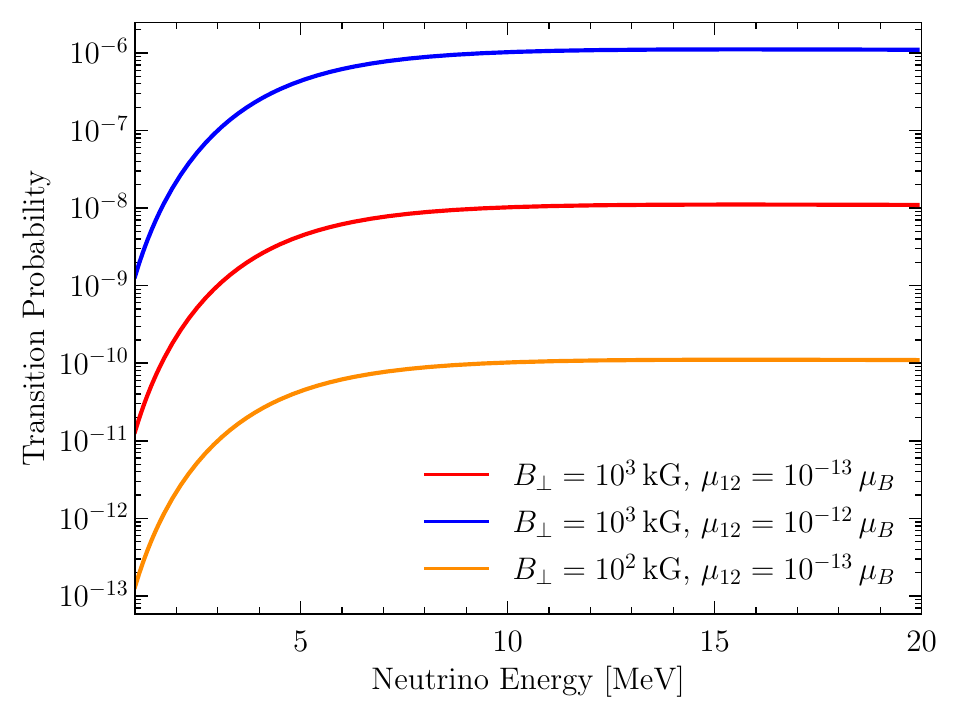}
    \caption{Transition probabilities in the solar neutrino energy $E_\nu$.
    }
    \label{fig:TransE}
\end{figure}

Recalling that the energy threshold of IBD is about 1.8 MeV~\cite{Cowan:1956rrn}, the observable solar neutrinos are $^8$B and $hep$ neutrinos. 
Due to the huge difference in flux, we practically focus on the $^8$B neutrinos as the source of the solar antineutrino appearance.
Following the Letter of Intent (LOI) of Jinping~\cite{Jinping:2016uua}, we take the solar neutrino energy spectra from Ref.~\cite{Bahcall},
normalized to the GS98 high metallicity prediction~\cite{Serenelli:2011py},$5.58\times 10^{6}\,$s$^{-1}$cm$^{-2}$. 
Then we adopt the approximate analytic expression of the transition probability $P(\nu_e \to \bar\nu_e)$ in the full 3-active flavor framework in Ref.~\cite{Akhmedov:2022txm}, which shows good agreement with full numerical calculations for neutrino energies above $\gtrsim 6$ MeV.
Figure~\ref{fig:TransE} displays the transition probabilities as a function of neutrino energy for representative fixed values of $\mu_{12}$ and $B_\perp$.
Interestingly, the probability for $E_\nu \gtrsim 8$ MeV can be approximated further as
\begin{align}
P(\nu_e \to \bar \nu_e) \simeq 1.1 \times 10^{-10} \left( \frac{\mu_{12} B_\perp}{10^{-12}\,\mu_B \cdot 10\,{\rm kG}}
\right)^2\,,
\label{eq:Psimpleflat}
\end{align}
which is asymptotically energy independent.
For $E_\nu \lesssim 6$ MeV, the full numerical results differ from the approximate analytic expression by a factor of few~\cite{Akhmedov:2022txm}.
Nevertheless, the contributions of such low energy neutrinos to the sensitivity analysis are relatively small due to the background contamination, as will be explained later.

The Jinping experiment is planned to use 4\,kt liquid scintillator or water-based liquid scintillator, with a fiducial 
mass of 3\,kt for IBD events.
In the low-energy range of $E_\nu\sim \mathcal{O} (1\, {\rm MeV})$, the IBD cross section can be approximated as~\cite{Giunti:2007ry}
\begin{align}
    \sigma_{\rm IBD}
    \approx
    1.601\times10^{-44}(1+3g_A^2)\left(\frac{E_\nu}{{\rm MeV}}\right)^2\,{\rm cm}^2\,,
\end{align}
where $g_A\approx 1.27$ is the axial vector coupling.

Combining the above processes together, we can 
express the predicted event rate of $\nu_e\rightarrow\bar\nu_e$ transition channel as
\begin{align}
    \frac{dN}{dE_\nu}
    =
    N_p  T \frac{d\phi_{\nu_e}}{dE_\nu}
    P(\nu_e\rightarrow\bar\nu_e)
    \sigma_{\rm IBD}\,,
\label{eq:dNdE}
\end{align}
where $N_p$ is the number of free protons inside the detector, $T$ is the data-taking time, and
$d\phi_{\nu_e}/dE_\nu$ is the solar neutrino energy spectrum. 
For IBD events with $\mathcal{O} (1)\, {\rm MeV}$ neutrinos, 
the kinetic energy of proton and neutron can be neglected. 
Hence the positron energy is related to the neutrino energy by $E_e\sim E_\nu - (m_n-m_p)$, where $m_n$ ($m_p$) is the neutron (proton) mass.
The produced positron will deposit its kinetic energy inside the detector and further
annihilate with an electron in the detector, 
emitting two 511\,keV photons. 
Consequently, the visible energy is contributed 
by the kinetic energy of the positron plus two annihilation photons,
\begin{align}
    E_{\rm vis}
    =
    E_e - m_e + 2 E_\gamma 
    \approx 
    E_\nu - 0.78\,{\rm MeV}.
\end{align}

\begin{figure}
    \centering    \includegraphics[width=0.5\linewidth]{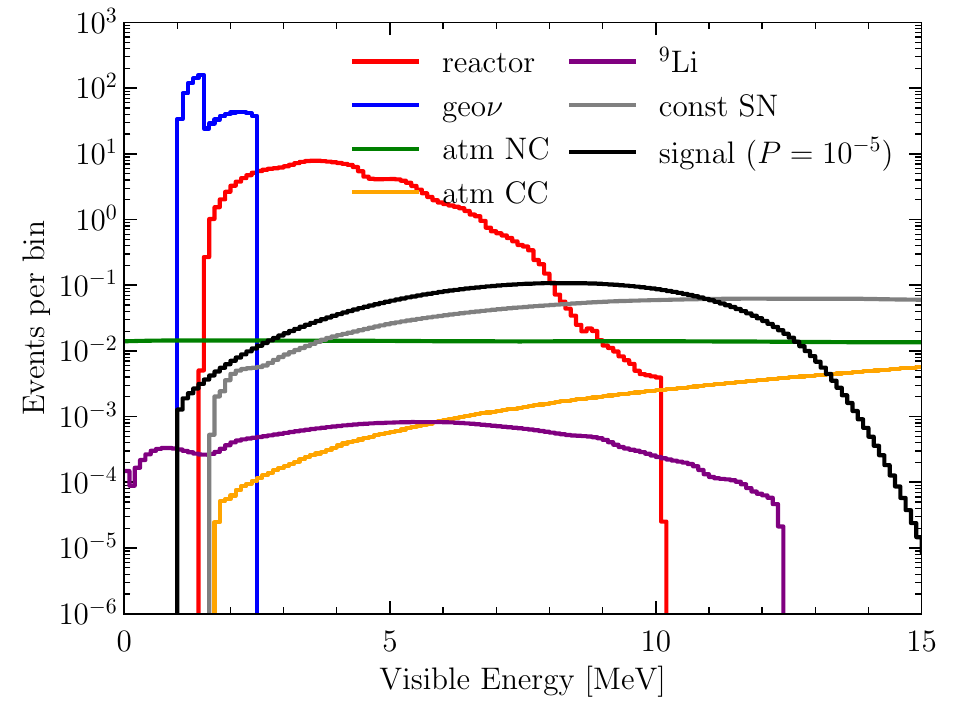}
    \caption{Differential spectrum of the expected number of signal (black) and the background events in Jinping for the 15 kt$\cdot$yr exposure.
    The horizontal axis is $E_{\rm vis} = E_\nu - 0.78$ MeV and the vertical axis is events per bin with the bin size of 0.1 MeV.
    }    \label{fig:JinpingEvents}
\end{figure}

The black solid curve in Fig.~\ref{fig:JinpingEvents} displays the 
differential IBD signal event rate as a function of visible energy, taking a benchmark transition probability $P(\nu_e \to \bar\nu_e)=10^{-5}$.
Here we take an exposure of $15\,{\rm kt}\cdot{\rm yr}$ with a bin width of $0.1$ MeV for illustration.
This event rate starts from $E_{\rm vis}\sim 1.1$ MeV, which corresponds to the 1.8 MeV energy threshold of IBD events. 
Beyond the energy threshold, the event rate increases up to $\lesssim 9$ MeV
since both neutrino flux and IBD cross section grow with the neutrino energy. 
However, for $E_{\rm vis}\gtrsim 9\,$MeV, the event rate quickly drops due to the sharp decrease of the neutrino flux.

We now discuss the six major backgrounds: the reactor neutrino, geoneutrino, atmospheric neutrino with NC and CC scattering, $^9$Li, and diffuse supernova neutrino background (DSNB).
The energy spectra of these backgrounds are extracted from \cite{Jinping:2016iiq} and shown in Fig.~\ref{fig:JinpingEvents}.
At low visible energies, $E_{\rm vis}\lesssim 8~{\rm MeV}$, the signal region is strongly contaminated by geoneutrino and cumulative reactor-antineutrino backgrounds, even though CJPL is located more than $1000~{\rm km}$ from nearby nuclear reactors. 
At higher visible energies, the dominant background is the DSNB contribution. 
For the DSNB flux, we adopt the constant-SN model of Ref.~\cite{Totani:1995rg}, which gives the largest flux for $E_\nu > 5~{\rm MeV}$ among the DSNB models considered in Ref.~\cite{Jinping:2016iiq}; this choice is therefore conservative. 
Although the DSNB has not yet been observed, we include it as an irreducible background in our analysis. 
In this model, the DSNB background becomes larger than the benchmark solar-$\bar\nu_e$ signal in the high-energy tail, $E_{\rm vis}\gtrsim 13~{\rm MeV}$.

In the realistic case,
the limited energy resolution of the detector,
which is quantified as $\delta E_{\rm vis}/E_{\rm vis}$,
will smear the true visible energy $E_{\rm vis}$
into the observed one, $E_{\rm vis}'$. Such an effect can 
be modeled by a Gaussian smearing function, $G(E_{\rm vis}, E_{\rm vis}')=\exp\left(-(E_{\rm vis}-E_{\rm vis}')^2/ 2(\delta E_{\rm vis})^2\right)/(\sqrt{2\pi}\delta E_{\rm vis})$.
In our study, we take an 8\% energy resolution for Jinping detector response \cite{Jinping:2016iiq}. 
To quantify the experimental sensitivity to the solar antineutrino appearance, we take the $\chi^2$ as
\begin{align}
    \chi^2
    \equiv
    2\sum_i^N\left[\mu_i - b_i + b_i \log\frac{b_i}{\mu_i}\right]
    +
    \frac{a^2}{\sigma_a^2}\,,\quad 
    {\rm with}\quad 
    \mu_i \equiv (1+a)s_i+b_i\,,
\end{align}
where $s_i$ ($b_i$) is the expected signal (background) event number in the $i$-th observed energy bin.
The binning size is taken as 0.1 MeV which is consistent with the energy spectra analyzed in \cite{Jinping:2016iiq}. 
For the systematics,
we assign an overall scaling nuisance parameter $a$ of 
the signal, with a $\sigma_a= 14\%$ normalization uncertainty \cite{Jinping:2016iiq}.
We then obtain the experimental constraint by minimizing $\chi^2$ over the nuisance parameter and subtracting the global
minimum value $\chi^2_{\rm min}$, $\Delta \chi^2\equiv \chi^2-\chi^2_{\rm min}$.

\begin{table}[h]
\centering
\begin{tabular}{llcc}
\hline
 &  & Low-background SR & Full SR \\
 &  & $E_{\rm vis}\in[8,13]~{\rm MeV}$ 
    & $E_{\rm vis}\in[1.02,16.02]~{\rm MeV}$ \\
 &  & $E_\nu\in[8.8,13.8]~{\rm MeV}$ 
    & $E_\nu\in[1.8,16.8]~{\rm MeV}$ \\
\hline
\multirow{7}{*}{Background~~~~~~~~~~~} 
 & Reactor $\bar\nu_e$      & 0.49 & 230.77 \\
 & Geo-$\bar\nu_e$          & 0.00 & 286.53 \\
 & Atmospheric CC           & 0.13 & 0.39 \\
 & Atmospheric NC           & 0.68 & 2.08 \\
 & $^9$Li                   & 0.01 & 0.05 \\
 & DSNB (const-SN model)    & 2.91 & 6.75 \\
 & Total background         & 4.22 & 526.57 \\
\hline
Signal 
 & $\nu_e\rightarrow \bar\nu_e$ 
 & 35.35 & 60.37 \\
\hline
\end{tabular}
\caption{Expected numbers of signal and background events at Jinping for an exposure of 
$3~{\rm kt}\times 5~{\rm yr}=15~{\rm kt}\cdot{\rm yr}$ in the two signal regions used in this analysis. 
The signal yield is evaluated for the benchmark values 
$B_\perp = 10^7\,{\rm G} = 10\,{\rm MG}$ and $\mu_{12}=10^{-12}\mu_B$. 
The DSNB background is evaluated using the constant-SN model of Ref.~\cite{Totani:1995rg}.
}
\label{tab:event}
\end{table}

To maximize the sensitivity, we focus mainly on the region $E_{\rm vis}\gtrsim 8~{\rm MeV}$, corresponding to $E_\nu \gtrsim 8.8~{\rm MeV}$, where the 
reactor-antineutrino background is strongly suppressed. 
This low-reactor-background environment suggests that the Jinping experiment can provide better sensitivity to solar $\bar\nu_e$ appearance than JUNO, despite having a fiducial mass roughly one seventh as large.
This corresponds to a low-background signal region (SR), $E_{\rm vis}\in[8,13]~{\rm MeV}$, corresponding to $E_\nu\in[8.8,13.8]~{\rm MeV}$.
For completeness, we also consider the full $^8$B-neutrino SR above the IBD threshold, $E_\nu\in[1.8,16.8]~{\rm MeV}$, corresponding to $E_{\rm vis}\in[1.02,16.02]~{\rm MeV}$.
We consider the latter region to facilitate a direct comparison with the JUNO sensitivity analysis in Ref.~\cite{Ventura:2025gfy}.
For $E_{\rm vis}\simeq 13~{\rm MeV}$, the expected signal yield is only about $0.4$ events for an exposure of $15~{\rm kt}\cdot{\rm yr}$ and is overwhelmed by the atmospheric-neutrino backgrounds as well as the DSNB. 
The expected signal and background event numbers in the two SRs are summarized in Table~\ref{tab:event}.

\subsection{Results}

\begin{figure}[h]
    \centering    \includegraphics[width=0.495\linewidth]{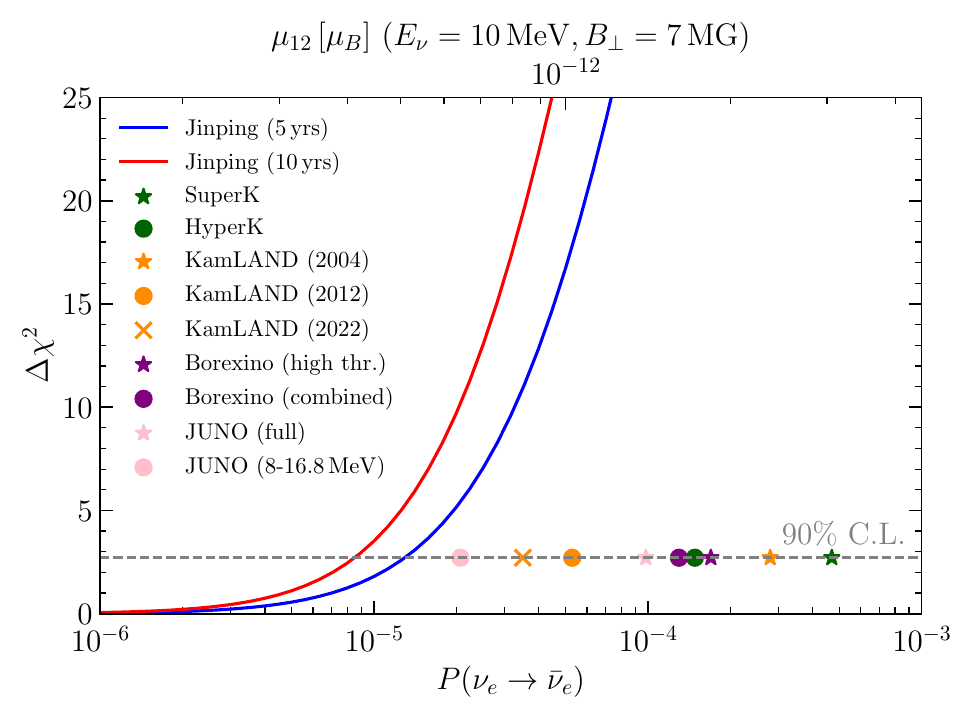}    \includegraphics[width=0.495\linewidth]{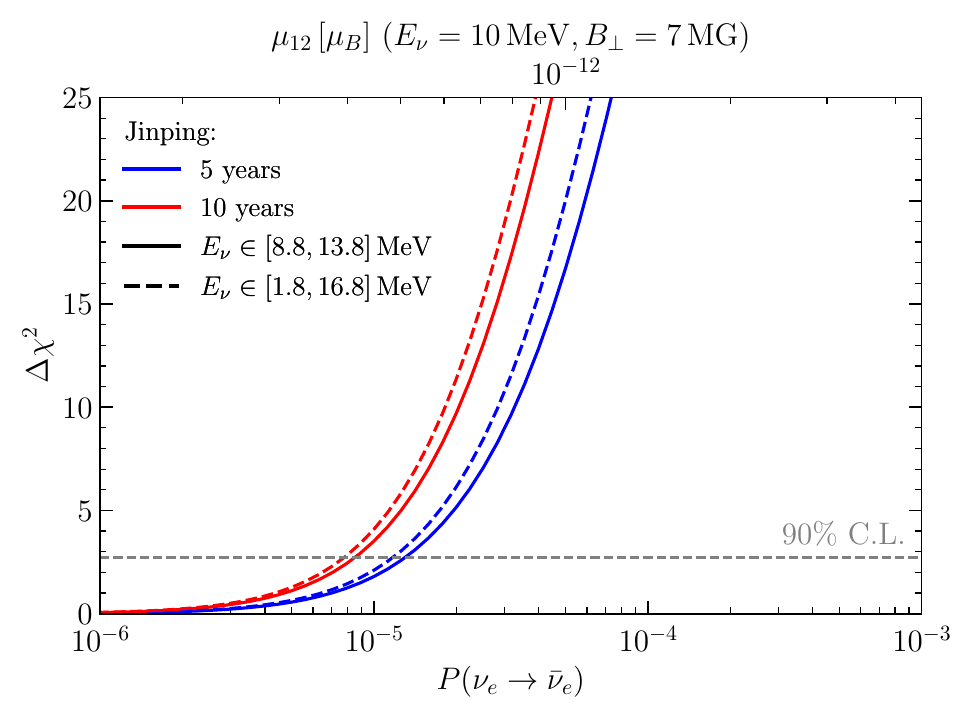}
    \caption{Sensitivities of Jinping to the transition probability $P(\nu_e \to \bar \nu_e)$ for the 5 (blue solid) and 10 years (red solid) of running with 3kt fiducial mass. 
    The horizontal dashed line corresponds to the 90\% C.L. sensitivity for a single degree of freedom.
    JUNO result is compared.
    }  \label{fig:JinpingTransition}
\end{figure}

Figure~\ref{fig:JinpingTransition} shows the results of the $\chi^2$ analysis described in the previous paragraph in the 
$\Delta\chi^2$--$P(\nu_e\to\bar\nu_e)$ plane. 
The projected sensitivities from the low-background SR for the 5 and 10 years of data taking at the Jinping experiment of a fiducial mass of $3~{\rm kt}$ are shown with blue and red solid lines, respectively. 
Note the transition probability $P(\nu_e\to\bar\nu_e)$ is dependent on neutrino energy as shown in Fig.~\ref{fig:TransE}. However, in this figure, we treat $P(\nu_e\to\bar\nu_e)$ of Eq.~(\ref{eq:dNdE}) as a free parameter which is energy-independent and then derive the experimental sensitivities to this parameter.
The horizontal dashed line denotes the 90\% C.L. criterion for one degree of freedom, corresponding to $\Delta\chi^2=2.71$. 
Owing to the low background rate in the low-background SR, the Jinping experiment can reach sensitivities at the level of 
$P(\nu_e\to\bar\nu_e)\sim 10^{-5}$. 
This is competitive with, and in some cases stronger than, the sensitivities expected from other experiments, including JUNO~\cite{Ventura:2025gfy}, despite the smaller fiducial mass of the Jinping experiment.

The upper horizontal axis shows the corresponding sensitivity to $\mu_{12}$ in the unit of $\mu_B$ for $E_\nu=10~{\rm MeV}$ and 
$B_\perp=7~{\rm MG}$ at $0.05 R_\odot$, dominantly determining the transition probability~\cite{Akhmedov:2022txm}.~\footnote{Actually the exact solar magnetic profile is unknown. Nevertheless, we follow the profile discussed in Ref.~\cite{Akhmedov:2022txm} normalizing $B_\perp (0.05 R_\odot) = 7$ MG or 10\,MG~\cite{Antia:2008}.}
This conversion is obtained from the approximate analytic expression in the 3-flavor neutrino framework in Ref.~\cite{Akhmedov:2022txm}. 
Under this benchmark assumption, the Jinping experiment can probe transition magnetic moments down to 
$\mu_{12}\sim (3\text{--}4)\times 10^{-13}\,\mu_B$.

For comparison, the left panel shows the existing constraints from KamLAND~\cite{KamLAND:2003gfh,KamLAND:2011bnd,KamLAND:2021gvi}, Borexino~\cite{Borexino:2010zht}, and Super-Kamiokande (SK)~\cite{Super-Kamiokande:2020frs}, together with the projected JUNO sensitivity for 5 years of data taking~\cite{Ventura:2025gfy}. 
The Borexino limits are shown separately for the high-threshold analysis, $E_\nu>7.3~{\rm MeV}$, which suppresses reactor-antineutrino backgrounds, and for the combined analysis over the full $^8$B region above the IBD threshold. 
For HK, we estimate the projected sensitivity by rescaling the SK constraint according to the exposure. 
The SK result is based on 2970.1 live days, corresponding to an exposure of approximately $183~{\rm kt}\cdot{\rm yr}$ for a fiducial mass of $22.5~{\rm kt}$, while the HK benchmark exposure is $188~{\rm kt}\times 9.74~{\rm yr}=1830.9~{\rm kt}\cdot{\rm yr}$. 
Thus, the HK exposure is larger by a factor of about 10, and we estimate the HK sensitivity as 
$P_{\rm HK}^{90} \simeq P_{\rm SK}^{90}/\sqrt{10}$, assuming statistics-dominated sensitivity.

In the right panel of Fig.~\ref{fig:JinpingTransition}, we compare the sensitivities obtained in the low-background SR, shown by solid lines, with those obtained in the full SR, shown by dashed lines. 
Although the full SR is more affected by background contamination, it contains approximately twice as many signal events. 
As a result, the full-SR analysis yields a slightly stronger sensitivity than the low-background-SR analysis.

\begin{figure}[h]
    \centering    \includegraphics[width=0.495\linewidth]{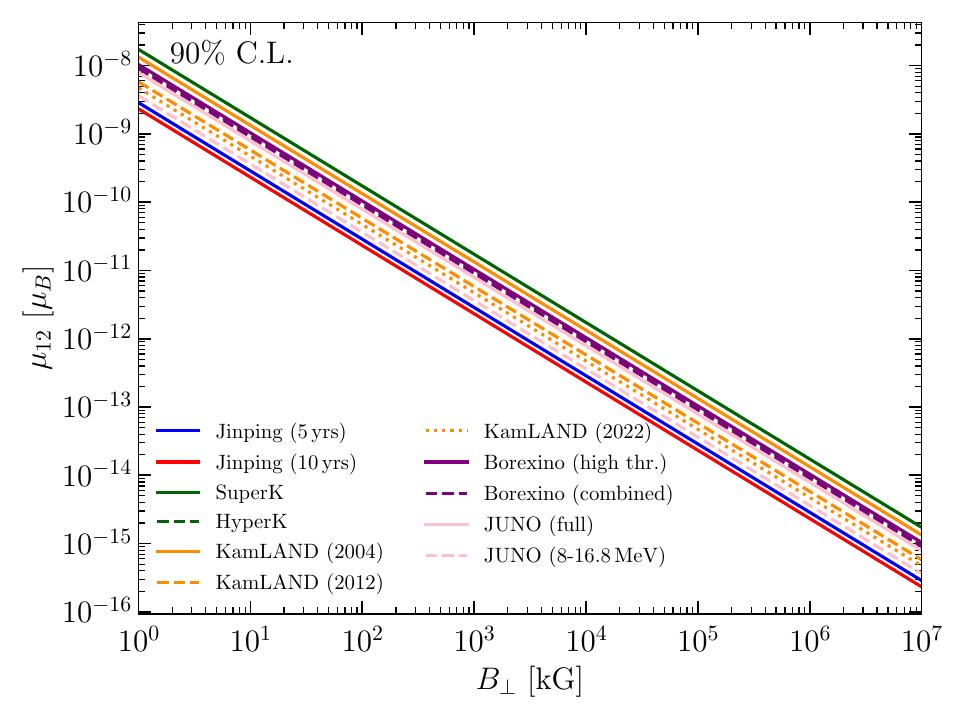}    \includegraphics[width=0.495\linewidth]{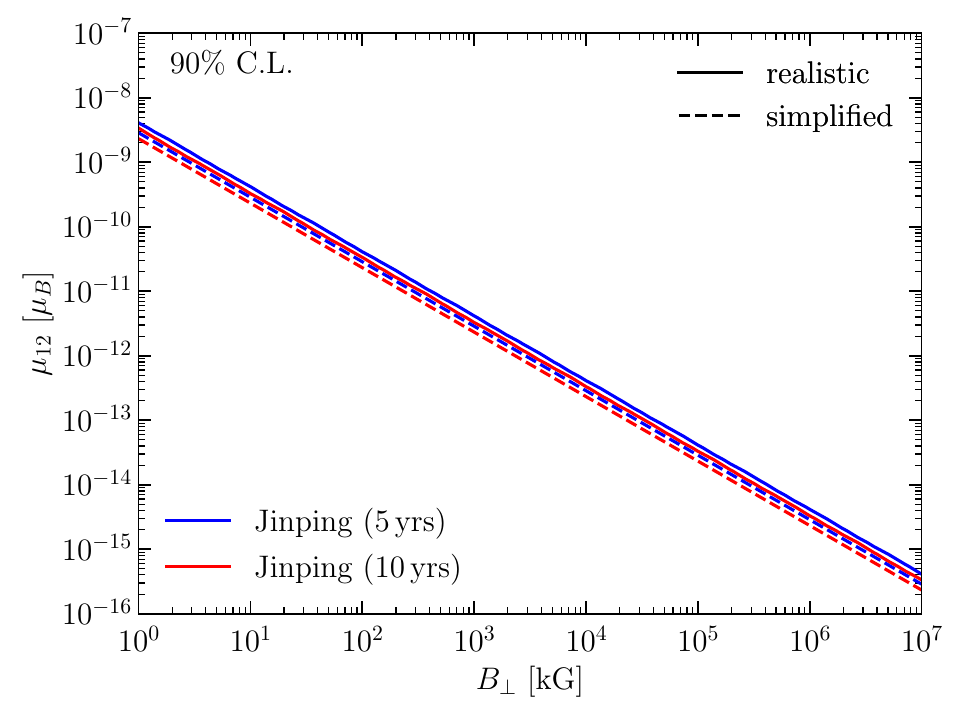}
    \caption{Sensitivities of Jinping to the neutrino transition magnetic moment for the 5 and 10 years of running with 3kt fiducial mass as a function of the transverse solar magnetic-field strength. Since the exact solar magnetic field profile is essentially unknown, we show the range up to $10^{10}$ G, although a helioseismic study suggests 7 MG in the solar core~\cite{Antia:2008}.
    }
    \label{fig:JinpingMoment}
\end{figure}

Since the strength of the transverse solar magnetic field, $B_\perp$, in the $^8$B-neutrino production region, 
$r\lesssim 0.05R_\odot$, is highly uncertain and model-dependent, we present the Jinping sensitivity to 
$\mu_{12}$ as a function of $B_\perp$ in Fig.~\ref{fig:JinpingMoment}. 
In this analysis, we scan a broad benchmark range, $B_\perp\in[1,10^7]~{\rm kG}$, using the asymptotically energy-independent simplified expression in Eq.~(\ref{eq:Psimpleflat}), which is valid for $E_\nu\gtrsim 8~{\rm MeV}$. 
The left panel compares the projected Jinping sensitivities with existing constraints and projected sensitivities from other experiments. 
The approximate linear behavior in the log-log plane reflects the fact that the transition probability is mainly controlled by the product $\mu_{12}B_\perp$.~\footnote{Note again that this is well justified for the analysis with our low-background SR.}

The right panel of Fig.~\ref{fig:JinpingMoment} compares the simplified Eq.~(\ref{eq:Psimpleflat}) and the approximate analytic formula of the transition probability in 
Ref.~\cite{Akhmedov:2022txm}, with dashed and solid lines, respectively.
We can see that our approach in obtaining the left panel of Fig.~\ref{fig:JinpingMoment} is good enough, although slightly more conservative sensitivity estimates are expected by using the approximate analytic formula.

\begin{table}[h]
\centering
\begin{tabular}{c|c|c|c}
 & $P(\nu_e\rightarrow\bar\nu_e)$
 & $\mu_{12}$ [$\mu_B$] ($B_\perp = 10\,$MG) &  $\mu_{12}$ [$\mu_B$] ($B_\perp = 7\,$MG)
\\
\hline
SK & $4.7\times 10^{-4}$ & $1.73 \times 10^{-12}$ & $2.47 \times 10^{-12}$
\\
HK & $1.5\times 10^{-4}$ & $9.73 \times 10^{-13}$ & $1.39 \times 10^{-12}$
\\
KamLAND 2004 & $2.8\times 10^{-4}$ & $1.34 \times 10^{-12}$ & $1.91 \times 10^{-12}$
\\
KamLAND 2012 & $5.3\times 10^{-5}$ & $5.81 \times 10^{-13}$ & $8.30 \times 10^{-13}$
\\ 
KamLAND 2022 & $3.5\times 10^{-5}$ & $4.72 \times 10^{-13}$ & $6.74 \times 10^{-13}$
\\
Borexino (high threshold) & $1.7\times 10^{-4}$ & $1.04 \times 10^{-12}$ & $1.49 \times 10^{-12}$
\\
Borexino (combined) & $1.3\times 10^{-4}$ & $9.10 \times 10^{-13}$ & $1.30 \times 10^{-12}$
\\
JUNO ($E_\nu\in [1.8, 16.8]\,$MeV, 5 year) & $9.8\times 10^{-5}$ & $7.92 \times 10^{-13}$ & $1.13 \times 10^{-12}$
\\
JUNO ($E_\nu\in [8, 16.8]\,$MeV, 5 year) & $2.1\times 10^{-5}$ & $3.63 \times 10^{-13}$ & $5.19 \times 10^{-13}$
\\
JUNO ($E_\nu\in [1.8, 16.8]\,$MeV, 10 year) & $6.9\times 10^{-5}$ & $6.63 \times 10^{-13}$ & $9.47 \times 10^{-13}$
\\
JUNO ($E_\nu\in [8, 16.8]\,$MeV, 10 year) & $1.5\times 10^{-5}$ & $3.09 \times 10^{-13}$ & $4.41 \times 10^{-13}$
\\
Jinping ($E_\nu\in [8.8, 13.8]\,$MeV, 5 year) & $1.3\times 10^{-5}$ & $2.87 \times 10^{-13}$ & $4.11 \times 10^{-13}$
\\
Jinping ($E_\nu\in [8.8, 13.8]\,$MeV, 10 year) & $8.5\times 10^{-6}$  & $2.33 \times 10^{-13}$ & $3.32 \times 10^{-13}$
\\
Jinping ($E_\nu\in [1.8, 16.8]\,$MeV, 5 year) & $1.2\times 10^{-5}$ & $2.73 \times 10^{-13}$ & $3.90 \times 10^{-13}$
\\
Jinping ($E_\nu\in [1.8, 16.8]\,$MeV, 10 year) & $7.8\times 10^{-6}$  & $2.22 \times 10^{-13}$ & $3.18 \times 10^{-13}$
\end{tabular}
\caption{Sensitivities/constraints on $P(\nu_e \to \bar \nu_e)$ and  $\mu_{12}$ for $B_\perp = 10\,$MG (3rd column) and 7\,MG (4th column) at $r = 0.05 R_\odot$.
}
\label{tab:constraint}
\end{table}

Published limits and projected sensitivities to the conversion probability $P(\nu_e\to\bar\nu_e)$ are summarized in the second column of Table~\ref{tab:constraint}. 
For the adopted $3\,\mathrm{kt}$ Jinping benchmark, we obtain projected 90\% C.L. sensitivities of $(1.2 - 1.3) \times 10^{-5}$ after five years and $(7.8 - 8.5)\times10^{-6}$ after ten years, depending on the signal region. 
These represent a factor-of-few improvement over the strongest existing bound listed in the table and approximately an order-of-magnitude improvement over the five-year JUNO projection for the full signal region. 
For the low-background SR, our five-year Jinping projection reaches a probability approximately $1.6$ times smaller than the corresponding five-year JUNO projection.

The third and fourth columns show the corresponding limits and projected sensitivities to $|\mu_{12}|$, inferred using the approximate analytic conversion formula in Ref.~\cite{Akhmedov:2022txm} for benchmark transverse magnetic fields of $B_\perp=10\,\mathrm{MG}$ and $7\,\mathrm{MG}$, respectively. 
Under these magnetic-field assumptions, both Jinping and JUNO with the restricted signal region could probe $|\mu_{12}|<10^{-12}\,\mu_B$.

For JUNO, we estimate the ten-year probability sensitivities by dividing the five-year results of Ref.~\cite{Ventura:2025gfy} by $\sqrt{2}$, assuming that the sensitivity remains dominated by background counting statistics and that systematic uncertainties are subdominant. 
The corresponding magnetic-moment sensitivities improve by a factor of $2^{1/4}$. 
These extrapolations and the comparison between experiments are conditional on the detector and background assumptions adopted in the respective analyses.

\begin{figure}[h]
    \centering    \includegraphics[width=\linewidth]{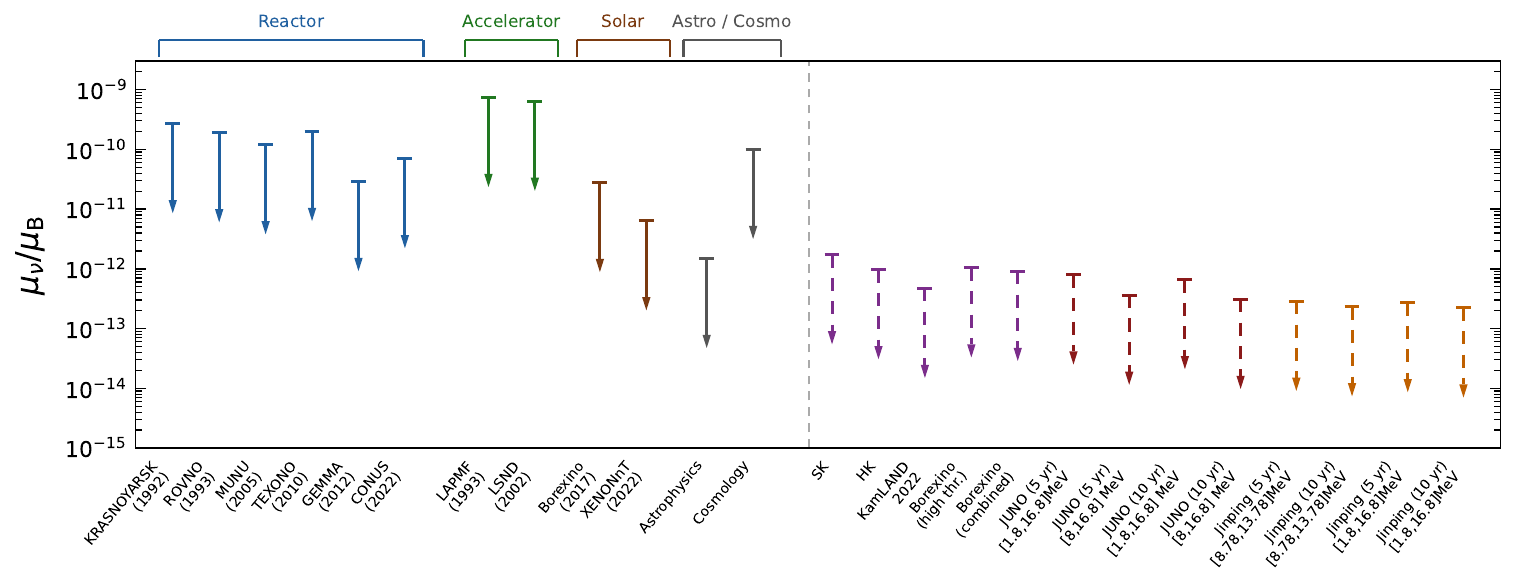}
    \caption{Summary of the sensitivities on the effective values of $\mu_\nu$. 
    }
    \label{fig:summary}
\end{figure}
In Fig.~\ref{fig:summary}, we present a comparison between the current experimental constraints on the $\nu$MDM (left panel) and the sensitivities that are achieved or expected in current and future antineutrino-appearance searches (right panel). Remarkably, our analysis indicates that forthcoming experiments could probe $\nu$MDM values as small as $(2\text{-}4)\times10^{-13}\,\mu_B$ depending on the magnitude of $B_\perp$ in the solar core. 
To the best of our knowledge, this would represent the most stringent projected sensitivity attainable in a laboratory experiment using established neutrino sources.~\footnote{A future nearby core-collapse supernova could provide an additional and potentially intriguing avenue for probing such small magnetic moments; see, e.g., Refs.~\cite{Jana:2023ufy, Jana:2022tsa, Ando:2002sk, Akhmedov:2003fu} for further discussion.} 
Notably, this projected reach would also surpass the most conservative astrophysical constraint, of order $10^{-12}\,\mu_B$, derived from observations of the luminosity at the TRGB~\cite{Capozzi:2020cbu}.

An important theoretical consideration is that Dirac neutrino magnetic moments larger than approximately $10^{-14}\,\mu_B$ are generally regarded as unnatural \cite{Bell:2005kz}, since radiative corrections can induce neutrino masses that are incompatible with their observed smallness. The sensitivity obtained in this work therefore approaches, and in fact extends beyond, the conventionally expected naturalness regime for Dirac neutrinos by roughly an order of magnitude. Consequently, an observation of antineutrino appearance at the sensitivity level discussed here in future experiments such as JUNO or Jinping would provide an intriguing indication in favor of Majorana neutrinos, provided the observed signal is established to originate from neutrino magnetic-moment effects. Such an interpretation follows from the expectation that Dirac magnetic moments of this magnitude would require additional protection against potentially large radiative contributions to the neutrino mass. Our results therefore motivate direct experimental searches for well-motivated BSM frameworks capable of accommodating comparatively large neutrino magnetic moments, including scenarios based on Voloshin symmetries, horizontal symmetries such as $SU(2)_{\rm H}$ and $SU(3)_{\rm H}$, or spin symmetries (see Refs.~\cite{Babu:2020ivd,Jana:2022xru,Babu:2021jnu} and references therein for further details).

\section{Conclusions}
\label{sec:conclusions}

In this work, we have revisited spin--flavor precession (SFP) of solar neutrinos induced by neutrino magnetic dipole moments ($\nu$MDMs) in the three-active-flavor framework, considering both Majorana and Dirac neutrinos. 
In the Majorana case, transition magnetic moments convert solar neutrinos into antineutrinos of different flavors. 
After propagation and flavor mixing, the resulting antineutrino state can contain a detectable $\bar{\nu}_e$ component, providing a distinctive lepton-number-violating solar-antineutrino appearance signal. 
In the Dirac case, SFP converts active left-handed neutrinos into right-handed states that are sterile with respect to the standard weak interactions, leading to active-neutrino disappearance and potentially to an energy-dependent distortion of the observed solar-neutrino spectrum.

We have analyzed the complete $6\times6$ Hamiltonians, including the vacuum kinetic term, the standard MSW matter potentials, and the magnetic interaction proportional to $\mu_\nu B_\perp(r)$. 
Because the magnetic interaction couples opposite-helicity sectors, a physical resonance cannot, in general, be identified simply by equating basis-dependent diagonal Hamiltonian elements. We instead identify candidate SFP resonances through crossings between propagation eigenvalues of the $B_\perp=0$ helicity blocks that are connected by a nonzero projected magnetic coupling. 
When the magnetic interaction is restored, such a crossing becomes an avoided crossing, and the corresponding finite-field resonance is associated with the magnetic-off propagation level crossing.

Using updated oscillation parameters and standard solar-model profiles based on the GS98 and AGSS09 compositions, we find no finite-density neutrino--antineutrino propagation-level crossing in the three-flavor Majorana case for the normal mass ordering and the neutrino-energy range $1\leq E_\nu/\mathrm{MeV}\leq20$ considered here. 
Majorana spin--flavor conversion in our setup is therefore nonresonant. Nevertheless, nonresonant SFP can still generate the distinctive solar $\bar{\nu}_e$ appearance signal.

In the Dirac case, by contrast, the propagation-level crossing $\lambda_1^L=d_1$ appears in the solar core for
\begin{align}
E_\nu \gtrsim 11.69~\mathrm{MeV}
\quad\text{(GS98)}\,,\qquad
E_\nu \gtrsim 12.21~\mathrm{MeV}
\quad\text{(AGSS09)}\,.
\end{align}
This crossing connects $\nu_{1m}^L$ with $\nu_{1R}$ rather than representing a pure $\nu_{eL}\leftrightarrow\nu_{eR}$ resonance. 
Moreover, the participating left-handed branch contains only a small electron-flavor component, and an outward-propagating neutrino encounters the crossing only if it is produced inside the corresponding resonance radius. 
These features strongly limit the resulting $\nu_e$ disappearance. 
For the transition-moment texture and strong core-field benchmarks adopted in this work, our numerical evolution nevertheless predicts a percent-level distortion of $P_{ee}$ in the upper $^8\mathrm{B}$ tail above approximately $12~\mathrm{MeV}$. 
Observing such a small distortion would be extremely challenging because of the limited endpoint statistics, uncertainties in the $^8\mathrm{B}$ spectral shape, the increasingly important and uncertain $hep$ contribution, and detector-response systematics. 
Future high-statistics measurements and dedicated spectral analyses may improve the sensitivity to this effect.

For the antineutrino appearance channel, we have assessed the prospective sensitivity of the proposed Jinping Neutrino Experiment, whose remote location provides a particularly low reactor-antineutrino background. 
Assuming a $3~\mathrm{kt}$ fiducial target mass for inverse beta decay, five or ten years of data taking, and optimistic benchmark transverse-field strengths of $B_\perp=7$ and $10~\mathrm{MG}$ in the solar core, we obtain a projected $90\%$ C.L. sensitivity to
\begin{align}
P(\nu_e\to\bar \nu_e)\simeq(0.8-1.3)\times10^{-5}.
\end{align}
Under these magnetic-field assumptions, the probability limits translate into a projected reach of
\begin{align}
|\mu_{12}|\simeq(2.2 - 4.1)\times10^{-13}\,\mu_B\,,
\end{align}
which is numerically below existing direct-scattering limits and commonly quoted stellar-cooling bounds. 
This inferred magnetic-moment reach is, however, conditional on the poorly known magnitude, geometry, and radial profile of the solar magnetic field, since the conversion probability is primarily controlled by the product $\mu_{12}B_\perp$.

Because our Jinping detector configuration and background estimates are based primarily on the 2016 Letter of Intent, these projections should be regarded as illustrative. 
An updated collaboration-level study incorporating the current detector design, efficiencies, energy response, and background estimates would therefore be valuable. 
JUNO and Hyper-Kamiokande may provide complementary solar-antineutrino searches through their different exposures and background environments, although our Hyper-Kamiokande estimate is based on an exposure rescaling rather than a complete detector-level analysis.

Finally, extending the present framework to include additional sterile states, such as in a $(3+1)$-flavor scenario, may introduce new active--sterile level crossings and energy-dependent disappearance signatures. 
A complete treatment of this possibility, including production averaging, enlarged-system evolution, and detector-level spectral analyses, is beyond the scope of the present work and is left for future study. 
Improved knowledge of the relevant solar-neutrino spectra, lower experimental thresholds, increased statistics, and better control of detector systematics will be important for exploring these extended scenarios.

\section*{Acknowledgments}
\label{sec:acknowledgments}
This work was initiated in the 8th workshop on Dark Matter as a Portal to New Physics (DMPNP 2026), the Focus Program [APCTP-2026-F01], held at APCTP, Pohang, Korea. 
The authors thank Evgeny K. Akhmedov, Andre De Gouvea, and Yago Porto for useful discussion.
The work of PB and SS is supported by the National Research Foundation of Korea (NRF) with Grant No. RS-2025-00562917.
SY is supported by Basic Science Research Program through NRF funded by the Ministry of Education (No. RS-2026-25571203).
CFK, SS, and SY are supported by the IBS fund IBS-R018-D1.

\bibliography{ref_solar_sfp}

@misc{Bahcall,
  author = {John N. Bahcall},
  title  = {John N. Bahcall Home Page},
  url    = {http://www.sns.ias.edu/~jnb/},
}

@article{JUNO:2015zny,
    author = "An, Fengpeng and others",
    collaboration = "JUNO",
    title = "{Neutrino Physics with JUNO}",
    eprint = "1507.05613",
    archivePrefix = "arXiv",
    primaryClass = "physics.ins-det",
    doi = "10.1088/0954-3899/43/3/030401",
    journal = "J. Phys. G",
    volume = "43",
    number = "3",
    pages = "030401",
    year = "2016"
}

@article{Jana:2022tsa,
    author = "Jana, Sudip and Porto-Silva, Yago P. and Sen, Manibrata",
    title = "{Exploiting a future galactic supernova to probe neutrino magnetic moments}",
    eprint = "2203.01950",
    archivePrefix = "arXiv",
    primaryClass = "hep-ph",
    doi = "10.1088/1475-7516/2022/09/079",
    journal = "JCAP",
    volume = "09",
    pages = "079",
    year = "2022"
}

@article{Ando:2002sk,
    author = "Ando, Shin'ichiro and Sato, Katsuhiko",
    title = "{Three generation study of neutrino spin flavor conversion in supernova and implication for neutrino magnetic moment}",
    eprint = "hep-ph/0211053",
    archivePrefix = "arXiv",
    reportNumber = "UTAP-419",
    doi = "10.1103/PhysRevD.67.023004",
    journal = "Phys. Rev. D",
    volume = "67",
    pages = "023004",
    year = "2003"
}

@article{Bell:2005kz,
    author = "Bell, Nicole F. and Cirigliano, Vincenzo and Ramsey-Musolf, Michael J. and Vogel, Petr and Wise, Mark B.",
    title = "{How magnetic is the Dirac neutrino?}",
    eprint = "hep-ph/0504134",
    archivePrefix = "arXiv",
    reportNumber = "CALT-08-2554, KRL-MAP-307",
    doi = "10.1103/PhysRevLett.95.151802",
    journal = "Phys. Rev. Lett.",
    volume = "95",
    pages = "151802",
    year = "2005"
}

@article{Akhmedov:2003fu,
    author = "Akhmedov, Evgeny K. and Fukuyama, Takeshi",
    title = "{Supernova prompt neutronization neutrinos and neutrino magnetic moments}",
    eprint = "hep-ph/0310119",
    archivePrefix = "arXiv",
    doi = "10.1088/1475-7516/2003/12/007",
    journal = "JCAP",
    volume = "12",
    pages = "007",
    year = "2003"
}

@article{Jana:2023ufy,
    author = "Jana, Sudip and Porto, Yago",
    title = "{Resonances of Supernova Neutrinos in Twisting Magnetic Fields}",
    eprint = "2303.13572",
    archivePrefix = "arXiv",
    primaryClass = "hep-ph",
    doi = "10.1103/PhysRevLett.132.101005",
    journal = "Phys. Rev. Lett.",
    volume = "132",
    number = "10",
    pages = "101005",
    year = "2024"
}

@article{Super-Kamiokande:2002weg,
    author = "Fukuda, Y. and others",
    editor = "Ilyin, V. A. and Korenkov, V. V. and Perret-Gallix, D.",
    collaboration = "Super-Kamiokande",
    title = "{The Super-Kamiokande detector}",
    doi = "10.1016/S0168-9002(03)00425-X",
    journal = "Nucl. Instrum. Meth. A",
    volume = "501",
    pages = "418--462",
    year = "2003"
}

@article{Hyper-Kamiokande:2018ofw,
    author = "Abe, K. and others",
    collaboration = "Hyper-Kamiokande",
    title = "{Hyper-Kamiokande Design Report}",
    eprint = "1805.04163",
    archivePrefix = "arXiv",
    primaryClass = "physics.ins-det",
    month = "5",
    year = "2018"
}

@article{Serenelli:2011py,
    author = "Serenelli, Aldo M. and Haxton, W. C. and Pena-Garay, Carlos",
    title = "{Solar models with accretion. I. Application to the solar abundance problem}",
    eprint = "1104.1639",
    archivePrefix = "arXiv",
    primaryClass = "astro-ph.SR",
    reportNumber = "UCB-NPAT-11-005, NT-LBNL-11-008",
    doi = "10.1088/0004-637X/743/1/24",
    journal = "Astrophys. J.",
    volume = "743",
    pages = "24",
    year = "2011"
}

@book{Giunti:2007ry,
    author = "Giunti, Carlo and Kim, Chung W.",
    title = "{Fundamentals of Neutrino Physics and Astrophysics}",
    doi = "10.1093/acprof:oso/9780198508717.001.0001",
    isbn = "978-0-19-850871-7",
    year = "2007"
}

@article{Cisneros:1970nq,
    author = "Cisneros, Arturo",
    title = "{Effect of neutrino magnetic moment on solar neutrino observations}",
    doi = "10.1007/BF00654607",
    journal = "Astrophys. Space Sci.",
    volume = "10",
    pages = "87--92",
    year = "1971"
}

@article{Okun:1986na,
    author = "Okun, L. B. and Voloshin, M. B. and Vysotsky, M. I.",
    title = "{Neutrino Electrodynamics and Possible Effects for Solar Neutrinos}",
    reportNumber = "ITEP-86-82",
    journal = "Sov. Phys. JETP",
    volume = "64",
    pages = "446--452",
    year = "1986"
}

@article{Mikheyev:1985zog,
    author = "Mikheyev, S. P. and Smirnov, A. Yu.",
    title = "{Resonance Amplification of Oscillations in Matter and Spectroscopy of Solar Neutrinos}",
    journal = "Sov. J. Nucl. Phys.",
    volume = "42",
    pages = "913--917",
    year = "1985"
}

@article{Giunti:2014ixa,
    author = "Giunti, Carlo and Studenikin, Alexander",
    title = "{Neutrino electromagnetic interactions: a window to new physics}",
    eprint = "1403.6344",
    archivePrefix = "arXiv",
    primaryClass = "hep-ph",
    doi = "10.1103/RevModPhys.87.531",
    journal = "Rev. Mod. Phys.",
    volume = "87",
    pages = "531",
    year = "2015"
}

@article{Akhmedov:2022txm,
    author = "Akhmedov, Evgeny and Mart{\'\i}nez-Mirav{\'e}, Pablo",
    title = "{Solar $ {\overline{\nu}}_e $ flux: revisiting bounds on neutrino magnetic moments and solar magnetic field}",
    eprint = "2207.04516",
    archivePrefix = "arXiv",
    primaryClass = "hep-ph",
    doi = "10.1007/JHEP10(2022)144",
    journal = "JHEP",
    volume = "10",
    pages = "144",
    year = "2022"
}

@article{Loureiro:2018pdz,
    author = "Loureiro, Arthur and others",
    title = "{On The Upper Bound of Neutrino Masses from Combined Cosmological Observations and Particle Physics Experiments}",
    eprint = "1811.02578",
    archivePrefix = "arXiv",
    primaryClass = "astro-ph.CO",
    doi = "10.1103/PhysRevLett.123.081301",
    journal = "Phys. Rev. Lett.",
    volume = "123",
    number = "8",
    pages = "081301",
    year = "2019"
}

@article{Jinping:2016iiq,
    author = "Beacom, John F. and others",
    collaboration = "Jinping",
    title = "{Physics prospects of the Jinping neutrino experiment}",
    eprint = "1602.01733",
    archivePrefix = "arXiv",
    primaryClass = "physics.ins-det",
    doi = "10.1088/1674-1137/41/2/023002",
    journal = "Chin. Phys. C",
    volume = "41",
    number = "2",
    pages = "023002",
    year = "2017"
}

@article{Ventura:2025gfy,
    author = "Ventura, C. V. and Churata, Saul J. Panibra",
    title = "{Estimating JUNO{\textquoteright}s Sensitivity to Solar Neutrino-to-antineutrino Conversion and Neutrino Magnetic Moments}",
    eprint = "2601.03502",
    archivePrefix = "arXiv",
    primaryClass = "hep-ph",
    doi = "10.3847/1538-4357/ae226d",
    journal = "Astrophys. J.",
    volume = "996",
    pages = "88",
    year = "2025"
}

@article{DeRomeri:2024hvc,
    author = "De Romeri, Valentina and Papoulias, Dimitrios K. and Sanchez Garcia, Gonzalo and Ternes, Christoph A. and T{\'o}rtola, Mariam",
    title = "{Neutrino electromagnetic properties and sterile dipole portal in light of the first solar CE{\ensuremath{\nu}}NS~data}",
    eprint = "2412.14991",
    archivePrefix = "arXiv",
    primaryClass = "hep-ph",
    doi = "10.1088/1475-7516/2025/05/080",
    journal = "JCAP",
    volume = "05",
    pages = "080",
    year = "2025"
}

@article{XENON:2024ijk,
    author = "Aprile, Elena and others",
    collaboration = "XENON",
    title = "{First Indication of Solar B8 Neutrinos via Coherent Elastic Neutrino-Nucleus Scattering with XENONnT}",
    eprint = "2408.02877",
    archivePrefix = "arXiv",
    primaryClass = "hep-ex",
    doi = "10.1103/PhysRevLett.133.191002",
    journal = "Phys. Rev. Lett.",
    volume = "133",
    number = "19",
    pages = "191002",
    year = "2024"
}

@article{PandaX:2024muv,
    author = "Bo, Zihao and others",
    collaboration = "PandaX",
    title = "{First Indication of Solar B8 Neutrinos through Coherent Elastic Neutrino-Nucleus Scattering in PandaX-4T}",
    eprint = "2407.10892",
    archivePrefix = "arXiv",
    primaryClass = "hep-ex",
    doi = "10.1103/PhysRevLett.133.191001",
    journal = "Phys. Rev. Lett.",
    volume = "133",
    number = "19",
    pages = "191001",
    year = "2024"
}

@article{XENON:2022ltv,
    author = "Aprile, E. and others",
    collaboration = "XENON",
    title = "{Search for New Physics in Electronic Recoil Data from XENONnT}",
    eprint = "2207.11330",
    archivePrefix = "arXiv",
    primaryClass = "hep-ex",
    doi = "10.1103/PhysRevLett.129.161805",
    journal = "Phys. Rev. Lett.",
    volume = "129",
    number = "16",
    pages = "161805",
    year = "2022"
}

@article{Capozzi:2020cbu,
    author = "Capozzi, Francesco and Raffelt, Georg",
    title = "{Axion and neutrino bounds improved with new calibrations of the tip of the red-giant branch using geometric distance determinations}",
    eprint = "2007.03694",
    archivePrefix = "arXiv",
    primaryClass = "astro-ph.SR",
    reportNumber = "MPP-2020-106",
    doi = "10.1103/PhysRevD.102.083007",
    journal = "Phys. Rev. D",
    volume = "102",
    number = "8",
    pages = "083007",
    year = "2020"
}

@article{Beda:2013mta,
    author = "Beda, A. G. and Brudanin, V. B. and Egorov, V. G. and Medvedev, D. V. and Pogosov, V. S. and Shevchik, E. A. and Shirchenko, M. V. and Starostin, A. S. and Zhitnikov, I. V.",
    title = "{Gemma experiment: The results of neutrino magnetic moment search}",
    doi = "10.1134/S1547477113020027",
    journal = "Phys. Part. Nucl. Lett.",
    volume = "10",
    pages = "139--143",
    year = "2013"
}

@article{Li:2022dkc,
    author = "Li, Shao-Ping and Xu, Xun-Jie",
    title = "{Neutrino magnetic moments meet precision N$_{eff}$ measurements}",
    eprint = "2211.04669",
    archivePrefix = "arXiv",
    primaryClass = "hep-ph",
    doi = "10.1007/JHEP02(2023)085",
    journal = "JHEP",
    volume = "02",
    pages = "085",
    year = "2023"
}

@article{Super-Kamiokande:2020frs,
    author = "Abe, K. and others",
    collaboration = "Super-Kamiokande",
    title = "{Search for solar electron anti-neutrinos due to spin-flavor precession in the Sun with Super-Kamiokande-IV}",
    eprint = "2012.03807",
    archivePrefix = "arXiv",
    primaryClass = "hep-ex",
    doi = "10.1016/j.astropartphys.2022.102702",
    journal = "Astropart. Phys.",
    volume = "139",
    pages = "102702",
    year = "2022"
}

@article{KamLAND:2003gfh,
    author = "Eguchi, K. and others",
    collaboration = "KamLAND",
    title = "{A High sensitivity search for anti-nu(e)'s from the sun and other sources at KamLAND}",
    eprint = "hep-ex/0310047",
    archivePrefix = "arXiv",
    doi = "10.1103/PhysRevLett.92.071301",
    journal = "Phys. Rev. Lett.",
    volume = "92",
    pages = "071301",
    year = "2004"
}

@article{KamLAND:2011bnd,
    author = "Gando, A. and others",
    collaboration = "KamLAND",
    title = "{A study of extraterrestrial antineutrino sources with the KamLAND detector}",
    eprint = "1105.3516",
    archivePrefix = "arXiv",
    primaryClass = "astro-ph.HE",
    doi = "10.1088/0004-637X/745/2/193",
    journal = "Astrophys. J.",
    volume = "745",
    pages = "193",
    year = "2012"
}

@article{KamLAND:2021gvi,
    author = "Abe, S. and others",
    collaboration = "KamLAND",
    title = "{Limits on Astrophysical Antineutrinos with the KamLAND Experiment}",
    eprint = "2108.08527",
    archivePrefix = "arXiv",
    primaryClass = "astro-ph.HE",
    doi = "10.3847/1538-4357/ac32c1",
    journal = "Astrophys. J.",
    volume = "925",
    number = "1",
    pages = "14",
    year = "2022"
}

@article{Borexino:2010zht,
    author = "Bellini, G. and others",
    collaboration = "Borexino",
    title = "{Study of solar and other unknown anti-neutrino fluxes with Borexino at LNGS}",
    eprint = "1010.0029",
    archivePrefix = "arXiv",
    primaryClass = "hep-ex",
    doi = "10.1016/j.physletb.2010.12.030",
    journal = "Phys. Lett. B",
    volume = "696",
    pages = "191--196",
    year = "2011"
}

@article{Babu:2020ivd,
    author = "Babu, K. S. and Jana, Sudip and Lindner, Manfred",
    title = "{Large Neutrino Magnetic Moments in the Light of Recent Experiments}",
    eprint = "2007.04291",
    archivePrefix = "arXiv",
    primaryClass = "hep-ph",
    reportNumber = "OSU-HEP-20-07",
    doi = "10.1007/JHEP10(2020)040",
    journal = "JHEP",
    volume = "10",
    pages = "040",
    year = "2020"
}

@article{Giunti:2024gec,
    author = "Giunti, Carlo and Kouzakov, Konstantin and Li, Yu-Feng and Studenikin, Alexander",
    title = "{Neutrino Electromagnetic Properties}",
    eprint = "2411.03122",
    archivePrefix = "arXiv",
    primaryClass = "hep-ph",
    doi = "10.1146/annurev-nucl-102122-023242",
    journal = "Ann. Rev. Nucl. Part. Sci.",
    volume = "75",
    number = "1",
    pages = "1--33",
    year = "2025"
}

@article{KamLAND:2013rgu,
    author = "Gando, A. and others",
    collaboration = "KamLAND",
    title = "{Reactor On-Off Antineutrino Measurement with KamLAND}",
    eprint = "1303.4667",
    archivePrefix = "arXiv",
    primaryClass = "hep-ex",
    doi = "10.1103/PhysRevD.88.033001",
    journal = "Phys. Rev. D",
    volume = "88",
    number = "3",
    pages = "033001",
    year = "2013"
    }

@article{SNO:2001kpb,
    author = "Ahmad, Q. R. and others",
    collaboration = "SNO",
    title = "{Measurement of the rate of $\nu_e+d \to p+p+e^-$ interactions produced by $^8$B solar neutrinos at the Sudbury Neutrino Observatory}",
    eprint = "nucl-ex/0106015",
    archivePrefix = "arXiv",
    reportNumber = "UPR-0240E",
    doi = "10.1103/PhysRevLett.87.071301",
    journal = "Phys. Rev. Lett.",
    volume = "87",
    pages = "071301",
    year = "2001"
}

@article{SNO:2002tuh,
    author = "Ahmad, Q. R. and others",
    collaboration = "SNO",
    title = "{Direct evidence for neutrino flavor transformation from neutral current interactions in the Sudbury Neutrino Observatory}",
    eprint = "nucl-ex/0204008",
    archivePrefix = "arXiv",
    doi = "10.1103/PhysRevLett.89.011301",
    journal = "Phys. Rev. Lett.",
    volume = "89",
    pages = "011301",
    year = "2002"
}

@article{Super-Kamiokande:1998kpq,
    author = "Fukuda, Y. and others",
    collaboration = "Super-Kamiokande",
    title = "{Evidence for oscillation of atmospheric neutrinos}",
    eprint = "hep-ex/9807003",
    archivePrefix = "arXiv",
    reportNumber = "BU-98-17, ICRR-REPORT-422-98-18, UCI-98-8, KEK-PREPRINT-98-95, LSU-HEPA-5-98, UMD-98-003, SBHEP-98-5, TKU-PAP-98-06, TIT-HPE-98-09",
    doi = "10.1103/PhysRevLett.81.1562",
    journal = "Phys. Rev. Lett.",
    volume = "81",
    pages = "1562--1567",
    year = "1998"
}

@article{Super-Kamiokande:2001ljr,
    author = "Fukuda, S. and others",
    collaboration = "Super-Kamiokande",
    title = "{Solar B-8 and hep neutrino measurements from 1258 days of Super-Kamiokande data}",
    eprint = "hep-ex/0103032",
    archivePrefix = "arXiv",
    doi = "10.1103/PhysRevLett.86.5651",
    journal = "Phys. Rev. Lett.",
    volume = "86",
    pages = "5651--5655",
    year = "2001"
}

@article{Akhmedov:1991nt,
    author = "Akhmedov, Evgeny K.",
    title = "{Oscillations - assisted resonant spin - flavor precession and time variations of the solar neutrino flux}",
    doi = "10.1016/0370-2693(91)90875-Q",
    journal = "Phys. Lett. B",
    volume = "257",
    pages = "163--167",
    year = "1991"
}

@article{Raghavan:1991em,
    author = "Raghavan, R. S. and Balantekin, A. B. and Loreti, F. and Baltz, A. J. and Pakvasa, S. and Pantaleone, James T.",
    title = "{Direct tests for solar neurino mass, mixing and majorana magnetic moment}",
    reportNumber = "UH-511-720-91",
    doi = "10.1103/PhysRevD.44.3786",
    journal = "Phys. Rev. D",
    volume = "44",
    pages = "3786--3790",
    year = "1991"
}

@article{Akhmedov:1987nc,
    author = "Akhmedov, Evgeny K.",
    title = "{Resonance enhancement of the neutrino spin precession in matter and the solar neutrino problem}",
    journal = "Sov. J. Nucl. Phys.",
    volume = "48",
    pages = "382--383",
    year = "1988"
}

@article{Antia:2008,
    author = "Antia, H.  M.",
    title = "{Seismic study of magnetic field in the solar interior}",
    journal = "J. Astrophys. Astron.",
    volume = "29",
    pages = "85–92",
    year = "2008"
}

@article{Kitchatinov:2008,
    author = "Kitchatinov, L.  L.",
    title = "{Stability of toroidal magnetic fields in the radiation zone of a star}",
    journal = "Astron. Rep.",
    volume = "52",
    pages = "247-255",
    year = "2008"

}

@article{Totani:1995rg,
    author = "Totani, Tomonori and Sato, Katsuhiko",
    title = "{Spectrum of the relic neutrino background from past supernovae and cosmological models}",
    eprint = "astro-ph/9504015",
    archivePrefix = "arXiv",
    reportNumber = "UTAP-203-95",
    doi = "10.1016/0927-6505(95)00015-9",
    journal = "Astropart. Phys.",
    volume = "3",
    pages = "367--376",
    year = "1995"
}

@article{Fujikawa:1980yx,
    author = "Fujikawa, Kazuo and Shrock, Robert",
    title = "{The Magnetic Moment of a Massive Neutrino and Neutrino Spin Rotation}",
    reportNumber = "ITP-SB-80-38",
    doi = "10.1103/PhysRevLett.45.963",
    journal = "Phys. Rev. Lett.",
    volume = "45",
    pages = "963",
    year = "1980"
}

@article{Vassh:2015yza,
    author = "Vassh, N. and Grohs, E. and Balantekin, A. B. and Fuller, G. M.",
    title = "{Majorana Neutrino Magnetic Moment and Neutrino Decoupling in Big Bang Nucleosynthesis}",
    eprint = "1510.00428",
    archivePrefix = "arXiv",
    primaryClass = "astro-ph.CO",
    doi = "10.1103/PhysRevD.92.125020",
    journal = "Phys. Rev. D",
    volume = "92",
    number = "12",
    pages = "125020",
    year = "2015"
}

@article{CONUS:2022qbb,
    author = "Bonet, H. and others",
    collaboration = "CONUS",
    title = "{First upper limits on neutrino electromagnetic properties from the CONUS experiment}",
    eprint = "2201.12257",
    archivePrefix = "arXiv",
    primaryClass = "hep-ex",
    doi = "10.1140/epjc/s10052-022-10722-1",
    journal = "Eur. Phys. J. C",
    volume = "82",
    number = "9",
    pages = "813",
    year = "2022"
}

@article{Voloshin:1986ty,
    author = "Voloshin, M. B. and Vysotsky, M. I. and Okun, L. B.",
    title = "{Neutrino electrodynamics and possible effects for solar neutrinos}",
    journal = "Sov. Phys. JETP",
    volume = "64",
    pages = "446",
    year = "1986"
}

@article{Lim:1987tk,
    author = "Lim, Chong-Sa and Marciano, William J.",
    title = "{Resonant spin-flavor precession of solar and supernova neutrinos}",
    doi = "10.1103/PhysRevD.37.1368",
    journal = "Phys. Rev. D",
    volume = "37",
    pages = "1368--1373",
    year = "1988"
}

@article{Akhmedov:1988uk,
    author = "Akhmedov, E. K.",
    title = "{Resonant amplification of neutrino spin rotation in matter and the solar-neutrino problem}",
    doi = "10.1016/0370-2693(88)91048-9",
    journal = "Phys. Lett. B",
    volume = "213",
    pages = "64--68",
    year = "1988"
}

@article{Wolfenstein:1977ue,
    author = "Wolfenstein, Lincoln",
    title = "{Neutrino oscillations in matter}",
    doi = "10.1103/PhysRevD.17.2369",
    journal = "Phys. Rev. D",
    volume = "17",
    pages = "2369--2374",
    year = "1978"
}

@article{Mikheev:1986gs,
    author = "Mikheev, S. P. and Smirnov, A. Yu.",
    title = "{Resonant amplification of neutrino oscillations in matter and solar-neutrino spectroscopy}",
    doi = "10.1007/BF02508049",
    journal = "Nuovo Cim. C",
    volume = "9",
    pages = "17--26",
    year = "1986"
}

@article{Jinping:2016uua,
    author = "Beacom, John F. and others",
    title = "{Letter of Intent: Jinping Neutrino Experiment}",
    eprint = "1602.01733",
    archivePrefix = "arXiv",
    primaryClass = "physics.ins-det",
    journal = "Chin. Phys. C",
    volume = "41",
    number = "2",
    pages = "023002",
    year = "2017"
}

@article{Picariello:2007qj,
    author = "Picariello, Marco and Pulido, Joao and Andringa, S. and Barros, N. F. and Maneira, J.",
    title = "{SNO+: predictions from standard solar models and resonant spin flavour precession}",
    eprint = "0705.4070",
    archivePrefix = "arXiv",
    primaryClass = "hep-ph",
    year = "2007"
}

@article{Friedland:2005xh,
    author = "Friedland, Alexander",
    title = "{Do solar neutrinos probe neutrino electromagnetic properties?}",
    eprint = "hep-ph/0505165",
    archivePrefix = "arXiv",
    reportNumber = "LA-UR-05-3141",
    month = "5",
    year = "2005"
}

@article{Akhmedov:1993sh,
    author = "Akhmedov, Evgeny K. and Petcov, S. T. and Smirnov, A. Yu.",
    title = "{Neutrinos with mixing in twisting magnetic fields}",
    eprint = "hep-ph/9301211",
    archivePrefix = "arXiv",
    reportNumber = "SISSA-170-92-EP",
    doi = "10.1103/PhysRevD.48.2167",
    journal = "Phys. Rev. D",
    volume = "48",
    pages = "2167--2181",
    year = "1993"
}

@article{Das:2009kw,
    author = "Das, C. R. and Pulido, Joao and Picariello, Marco",
    title = "{Light sterile neutrinos, spin flavour precession and the solar neutrino experiments}",
    eprint = "0902.1310",
    archivePrefix = "arXiv",
    primaryClass = "hep-ph",
    doi = "10.1103/PhysRevD.79.073010",
    journal = "Phys. Rev. D",
    volume = "79",
    pages = "073010",
    year = "2009"
}

@article{Cowan:1956rrn,
    author = "Cowan, C. L. and Reines, F. and Harrison, F. B. and Kruse, H. W. and McGuire, A. D.",
    title = "{Detection of the free neutrino: A Confirmation}",
    doi = "10.1126/science.124.3212.103",
    journal = "Science",
    volume = "124",
    pages = "103--104",
    year = "1956"
}

@article{Jana:2022xru,
    author = "Jana, Sudip",
    title = "{Horizontal Symmetry and Large Neutrino Magnetic Moments}",
    doi = "10.22323/1.405.0037",
    journal = "PoS",
    volume = "DISCRETE2020-2021",
    pages = "037",
    year = "2022"
}

@article{Babu:2021jnu,
    author = "Babu, K. S. and Jana, Sudip and Lindner, Manfred and K, Vishnu P.",
    title = "{Muon g {\ensuremath{-}} 2 anomaly and neutrino magnetic moments}",
    eprint = "2104.03291",
    archivePrefix = "arXiv",
    primaryClass = "hep-ph",
    doi = "10.1007/JHEP10(2021)240",
    journal = "JHEP",
    volume = "10",
    pages = "240",
    year = "2021"
}

@article{Joshi:2019dcj,
    author = "Joshi, Sandeep and Jain, Sudhir R.",
    title = "{Neutrino spin-flavor oscillations in solar environment}",
    eprint = "1906.09475",
    archivePrefix = "arXiv",
    primaryClass = "hep-ph",
    doi = "10.1088/1674-4527/20/8/123",
    journal = "Res. Astron. Astrophys.",
    volume = "20",
    number = "8",
    pages = "123",
    year = "2020"
}

@article{Esteban:2024eli,
    author = "Esteban, Ivan and Gonzalez-Garcia, M. C. and Maltoni, Michele and Martinez-Soler, Ivan and Pinheiro, Jo{\~a}o Paulo and Schwetz, Thomas",
    title = "{NuFit-6.0: updated global analysis of three-flavor neutrino oscillations}",
    eprint = "2410.05380",
    archivePrefix = "arXiv",
    primaryClass = "hep-ph",
    reportNumber = "IFT-UAM/CSIC-24-140, YITP-SB-2024-24, IPPP/24/64, IPPP/24/64, IFT-UAM/CSIC-24-140, YITP-SB-2024-24",
    doi = "10.1007/JHEP12(2024)216",
    journal = "JHEP",
    volume = "12",
    pages = "216",
    year = "2024"
}

@article{Bakhti:2023vzn,
    author = "Bakhti, Pouya and Rajaee, Meshkat and Seo, Seon-Hee and Shin, Seodong",
    title = "{Exploring solar neutrino oscillation parameters with the liquid scintillator counter at Yemilab with a comparison to JUNO}",
    eprint = "2307.11582",
    archivePrefix = "arXiv",
    primaryClass = "hep-ph",
    reportNumber = "FERMILAB-PUB-23-746-PPD",
    doi = "10.1103/PhysRevD.109.095030",
    journal = "Phys. Rev. D",
    volume = "109",
    number = "9",
    pages = "095030",
    year = "2024"
}

@article{NuEYE:2026gyx,
    author = "CHEN, Shaomin and others",
    collaboration = "NuEYE",
    title = "{The $ν$EYE Neutrino Telescope: Conceptual Design Report}",
    eprint = "2601.12569",
    archivePrefix = "arXiv",
    primaryClass = "hep-ex",
    month = "1",
    year = "2026"
}

@article{Seo:2023xku,
    author = "Seo, Seon-Hee and others",
    title = "{Physics Potential of a Few Kiloton Scale Neutrino Detector at a Deep Underground Lab in Korea}",
    eprint = "2309.13435",
    archivePrefix = "arXiv",
    primaryClass = "hep-ex",
    month = "9",
    year = "2023"
}
\end{document}